\documentclass[sigconf,screen,pbalance=true]{acmart}

\usepackage{tikz}
\usepackage{amsmath}
\usepackage{graphicx}
\usepackage[many]{tcolorbox}

\usepackage{amsmath,amssymb,amsfonts}
\usepackage{textcomp}
\usepackage{xcolor,soul}
\usepackage[normalem]{ulem}
\usepackage{amsmath}
\usepackage{mathtools,amssymb,latexsym,amsfonts,stmaryrd}
\usepackage{pbox}
\usepackage{xfrac}
\usepackage{booktabs}
\usepackage{xcolor}
\usepackage{threeparttable}
\usepackage{amsthm}
\usepackage{multirow}
\usepackage{tikz}
\usepackage{mathrsfs}
\usepackage{mathpartir}
\usepackage{algorithm}
\usepackage{url}
\usepackage{syntax}
\usepackage{framed}
\usepackage{algpseudocode}
\usepackage{flushend}
\usepackage{listings}
\usepackage{moresize}
\usepackage{wrapfig}
\usepackage{xcolor}
\usepackage{colortbl}
 
\usepackage{seqsplit}
\usepackage{alltt}
\usepackage{xspace}
\usepackage{makecell}
\usepackage{subcaption}
\usepackage[toc,page]{appendix}
\usepackage{diagbox}
\usepackage{rotating}

\usepackage{pifont}

\DeclareMathAlphabet{\mathcal}{OMS}{cmsy}{m}{n}

\usepackage{thmtools}

\declaretheoremstyle[spaceabove=\topsep,notefont=\normalfont\itshape]{mystyle}

\newcommand{\revise}[2]{{\color{red}{\ifx&#1&\else- #1\fi}} {\color{ForestGreen}{\ifx&#2&\else+ #2\fi}}}\renewcommand{\revise}[2]{#2}

\usetikzlibrary{arrows,patterns,decorations.pathreplacing}

\usepackage{microtype}

\newcommand{\F}{Fig.}
\newcommand{\E}{Eq.}
\newcommand{\T}{Table}

\newcommand{\ignore}[1]{}

\lstdefinestyle{base}{
  moredelim=**[is][\color{red}]{@}{@},
  escapeinside={<@}{@>}
}

\newcommand{\parh}[1]{\noindent\textbf{#1}}

\usepackage{amssymb}\usepackage{pifont}

\newcommand\DejaVuttfamily{\fontfamily{DejaVuSansMono-TLF}\selectfont }

\lstdefinestyle{base}{
  moredelim=**[is][\color{red}]{@}{@},
  escapeinside={<@}{@>}
}

\lstdefinelanguage
   [x64]{Assembler}     [x86masm]{Assembler} {morekeywords={CDQE,CQO,CMPSQ,CMPXCHG16B,JRCXZ,LODSQ,MOVSXD, POPFQ,PUSHFQ,SCASQ,STOSQ,IRETQ,RDTSCP,SWAPGS, rax,rdx,rcx,rbx,rsi,rdi,rsp,rbp, r8,r8d,r8w,r8b,r9,r9d,r9w,r9b}} 

\usepackage{listings}
\usepackage{fancyvrb}
\usepackage{color}
\definecolor{lightgray}{rgb}{.9,.9,.9}
\definecolor{darkgray}{rgb}{.4,.4,.4}
\definecolor{purple}{rgb}{0.65, 0.12, 0.82}
\definecolor{commentgreen}{RGB}{63,127,95}

\colorlet{myPurple}{blue!40!red}
\definecolor{myOrange}{RGB}{255,192,0}

\lstdefinelanguage{Solidity}{
  keywords={len,delete,int,void,payable, public, event, contract, typeof, new, true, false, catch, function, return, null, catch, switch, var, if, in, while, do, else, case, break,struct,const,socklen_t,sa_familty_t,char,sockaddr},
  keywordstyle=\color{violet}\bfseries,
  ndkeywords={class, export, boolean, throw, implements, import, this},
  ndkeywordstyle=\color{darkgray}\bfseries,
  identifierstyle=\color{black},
  sensitive=false,
  comment=[l]{//},
  escapeinside={(*@}{@*)},          morecomment=[s]{/*}{*/},
  commentstyle=\color{commentgreen}\ttfamily,
  stringstyle=\color{red}\ttfamily,
  morestring=[b]',
  morestring=[b]"
}

\newcommand{\rnum}[1]{\uppercase\expandafter{\romannumeral #1\relax}}

\usepackage{xcolor}

\algnewcommand{\LeftComment}[1]{\Statex \(\triangleright\) #1}

\definecolor{pptbrown}{RGB}{132,60,12}
\definecolor{pptgreen}{RGB}{56,87,35}

\let\OLDthebibliography\thebibliography
\renewcommand\thebibliography[1]{
  \OLDthebibliography{#1}
  \setlength{\parskip}{0pt}
  \setlength{\itemsep}{0pt plus 0.1ex}
}

\definecolor{pptgreen}{RGB}{84,130,53}
\definecolor{pptred}{RGB}{176,35,24}
\definecolor{pptblue}{RGB}{194,214,236}

\definecolor{pptgreen1}{RGB}{78,173,91}
\definecolor{pptred1}{RGB}{192,0,0}

\definecolor{pptyellow1}{RGB}{203,195,167}
\definecolor{pptgreen2}{RGB}{184,192,176}

\newenvironment{packeditemize}{
\begin{list}{$\bullet$}{
\setlength{\labelwidth}{5pt}
\setlength{\itemsep}{0pt}
\setlength{\leftmargin}{\labelwidth}
\addtolength{\leftmargin}{\labelsep}
\setlength{\parindent}{0pt}
\setlength{\listparindent}{\parindent}
\setlength{\parsep}{0pt}
\setlength{\topsep}{0pt}}}{\end{list}}

 \newcommand{\mrrinsn}{MRR$^{\text{insn}}$\xspace}
\newcommand{\rinsn}{Recall$^{\text{insn}}$@1\xspace}

\newcommand{\rfunc}{Recall$^{\text{func}}$@1\xspace}

\newcommand{\tool}{\texttt{InsnAlign}\xspace}
\newcommand{\jtrans}{\texttt{jTrans}\xspace}
\newcommand{\jclap}{\texttt{CLAP}\xspace}

\newcommand{\toolj}{${\tool}_{\texttt{jtrans}}$\xspace}
\newcommand{\toolc}{${\tool}_{\texttt{clap}}$\xspace}

\definecolor{lightgreen}{HTML}{b4e5a2}
\definecolor{lightblue}{HTML}{c1e5f5}
\definecolor{lightorange}{HTML}{fbe3d6}
\definecolor{lightyellow}{HTML}{f5f5c1}
\definecolor{lightpurple}{HTML}{f2cfee}
\definecolor{plum}{HTML}{76c2d4}
\definecolor{gray}{HTML}{bfbfbf}

\title{Instruction Alignment for Binary Code Representation Learning}

\author{Huaijin Wang}
\authornote{Corresponding author}
\orcid{0000-0002-1066-0331}
\affiliation{\institution{Shandong University}
  \country{}
}
\email{huaijinwang@sdu.edu.cn}

\author{Shuai Wang}
\orcid{0000-0002-0866-0308}
\affiliation{\institution{Hong Kong University of Science and Technology}
  \country{}
}
\email{shuaiw@cse.ust.hk}

\copyrightyear{2026}
\acmYear{2026}
\setcopyright{cc}
\setcctype{by-nc-nd}
\acmConference[ASE '26]{Proceedings of the 41st IEEE/ACM International Conference on Automated Software Engineering}{October 12--16, 2026}{Munich, Germany}
\acmBooktitle{Proceedings of the 41st IEEE/ACM International Conference on Automated Software Engineering (ASE '26), October 12--16, 2026, Munich, Germany}
\acmDOI{10.1145/3832783.3837516}
\acmISBN{979-8-4007-2882-2/2026/10}

\begin{document}

\ccsdesc[500]{Security and privacy~Software reverse engineering}
\keywords{Representation Learning; Binary Code Similarity}

\begin{abstract}
Binary code representation learning is a fundamental problem in software security and reverse engineering. Existing methods mainly learn function-level embeddings that capture coarse-grained semantic relationships between binary functions, but they largely ignore fine-grained instruction-level correspondences. This limitation misses valuable supervision signals available from compiler debug information, which can support the learning of more accurate and interpretable binary code representations.

We propose to leverage instruction alignment knowledge to further improve binary code representation learning.
Our preliminary study reveals that models finetuned for function-level binary code similarity exhibit substantially better instruction alignment than their pre-trained model, suggesting a strong correlation between instruction alignment and function-level embedding quality.
Motivated by this observation, we design a training approach that explicitly incorporates instruction alignment as an auxiliary training objective.
Our experiments show that instruction alignment training improves retrieval accuracy and provides more discriminative signal for the model's similarity judgments.
\end{abstract}

\maketitle

\section{Introduction}
\label{sec:introduction}

Binary code representation learning trains models to map binary functions into fixed-dimensional vector embeddings. It has become a cornerstone of modern binary analysis~\cite{wang2024clap,kim2022revisiting,liu2022sok}.
High-quality binary code embeddings enable a wide range of downstream tasks, including vulnerability detection~\cite{ding2019asm2vec,zuo2019neural,luo2023vulhawk}, malware analysis~\cite{wong2022deceiving,massarelli2019safe,liu2025keenhash}, plagiarism detection~\cite{zuo2019neural,xu2023pem}, and software supply chain security~\cite{bsa2bsca2024,safesca2025,liu2025keenhash,jiang2024binaryai,yu2020codecmr}.
The embeddings' quality directly determines the effectiveness of downstream applications: embeddings that faithfully capture binary code semantics yield better retrieval, classification, and matching performance across the board.\looseness=-1

To train such embedding models, existing methods predominantly rely on function-level supervision~\cite{wang2017imf,haq2021survey,wang2022jtrans,kim2022revisiting,wang2024clap,marcelli2022machine,massarelli2019safe,ding2019asm2vec,xu2017neural,yu2020codecmr,yu2020order,sem2vec2023,wong2024binaug,wang2022enhancing,wang2026vsim}.
They compile the same source project under multiple configurations, such as various compilers (e.g., GCC and Clang) and optimization flags (e.g., \texttt{O0}--\texttt{O3}), and match binary functions across variants using their debug symbols.
Two binary functions sharing the same symbol form a positive pair, as they originate from the same source function; otherwise, they form a negative pair~\cite{xu2017neural,zuo2019neural,ding2019asm2vec,wang2022jtrans}.
Models are then trained with contrastive objectives, such as triplet loss, to produce embeddings that bring positive pairs close and push negative pairs apart in the embedding space.

While this function-level symbol matching has proven effective, it represents only the coarse-grained knowledge available from the compilation process.
Modern compilers produce a wealth of additional information. Most notably, debug information that maps each assembly instruction back to its originating source line.
This fine-grained mapping between source and binary code provides rich semantic knowledge that existing approaches have overlooked.\looseness=-1

This paper proposes to exploit \emph{instruction alignment} knowledge to improve binary code representation learning.
Specifically, we leverage the debug information produced during compilation to establish fine-grained correspondences between individual assembly instructions across different binary variants of the same function.
Two assembly instructions originating from the same source line are considered semantically aligned, providing a supervision signal at a much finer granularity than function-level symbol matching.\looseness=-1

We first conduct a measurement study to evaluate how well existing models capture instruction-level semantics.
We formulate instruction alignment as a retrieval task: given a pair of binary functions compiled from the same source, and given an instruction in one function, we measure the model's ability to retrieve the semantically corresponding instruction in the other function.
Our measurement reveals that models finetuned with function-level contrastive learning exhibit substantially better instruction alignment than their pre-trained version, suggesting a strong correlation between instruction-level understanding and embedding quality.

Motivated by this observation, we propose \tool, a training approach that explicitly incorporates instruction alignment as an auxiliary objective alongside function-level contrastive learning.
By augmenting the standard training pipeline with an InfoNCE loss~\cite{oord2018representation} on instruction-level correspondences, we improve embedding quality while providing interpretable, instruction-level evidence for similarity judgments.
The resulting model not only produces more accurate function-level embeddings but also provides interpretable evidence for similarity judgments by identifying which specific instructions correspond to each other.
Our contributions are summarized as follows:
\begin{packeditemize}
    \item We identify that existing binary code representation learning methods only exploit coarse-grained function-level knowledge and overlook the instruction-level semantic correspondences available from compiler debug information.\looseness=-1
    \item To our best knowledge, we are the first to propose an instruction alignment measurement to evaluate how well binary code embedding models capture fine-grained instruction semantics, formulated as a retrieval task for quantitative evaluation.
    \item We design a training approach that leverages instruction alignment knowledge to improve binary code embeddings.
    {Our extensive experiments demonstrate that our method improves retrieval performance while providing inspectable evidence that is more discriminative on hard-to-distinguish candidates.}
\end{packeditemize}
 \section{Background and Preliminaries}
\label{sec:background}

\subsection{ML-based BCSA}
\label{sec:background:bcsa}

Binary code similarity analysis (BCSA) is often adopted to measure the binary code representation quality. It aims to determine whether two binary code snippets share similar functionality.
This task is essential in scenarios where source code is unavailable, such as analyzing proprietary software~\cite{ming2017binsim}, detecting known vulnerabilities in firmware~\cite{jiang2024binaryai,liu2025keenhash}, and identifying code reuse across binaries~\cite{bsa2bsca2024}.
A common definition~\cite{wang2026vsim,wang2022jtrans,wong2024binaug,wang2024clap} of the BCSA task is as follows:

\smallskip
\parh{Definition.}
\textit{Given a binary function $q$ and a pool of candidate functions $P = \{f_1, f_2, \ldots, f_N\}$, the goal is to rank all candidates based on their similarity to $q$, ideally placing the true positive (i.e., the function compiled from the same source) at the top of the list.}

Because the pool of candidate functions can be very huge in real-world usage (e.g., millions of functions across firmware images or software repositories), fast similarity computation is critical~\cite{jiang2024binaryai}.
Conventional approaches that rely on symbolic execution~\cite{ming2017binsim,luo2017semantics,luo2014semantics,wang2022enhancing,chandramohan2016bingo} or dynamic testing~\cite{manuel2014blanket,wang2017imf} require heavyweight per-pair analysis, making them impractical at scale~\cite{ding2019asm2vec,wang2026vsim}.
In contrast, representation learning techniques support fast comparison inherently: the embedding model $\mathcal{M}$ maps a binary function $f$ to a fixed-dimensional vector $\mathbf{v} = \mathcal{M}(f) \in \mathbb{R}^d$, and similarity between any two functions reduces to a single vector distance computation (e.g., cosine similarity).
The sufficiently close embeddings indicate that the two functions are likely compiled from the same source function, while distant embeddings suggest dissimilarity.
This paradigm enables efficient retrieval over large candidate pools, as embeddings can be precomputed and indexed~\cite{jiang2024binaryai,yu2020codecmr,bsa2bsca2024}.
To train such models, a large dataset of binary functions with known semantic relationships is required~\cite{kim2022revisiting,wang2022jtrans}.

\begin{figure}
\centering
\includegraphics[width=\linewidth]{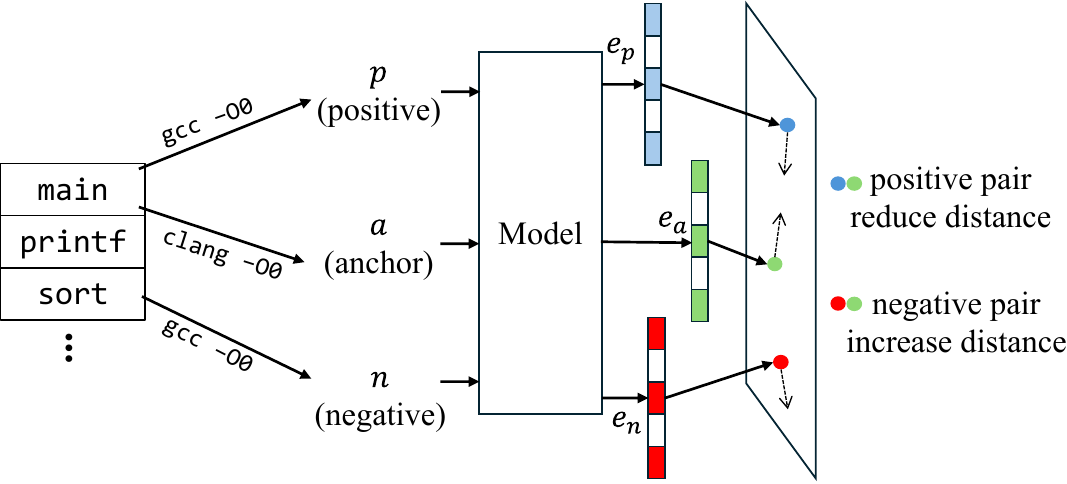}
\caption{Ground truth construction and training objective. $p$ and $a$ are compiled from ``\texttt{main}'' function, while $n$ is compiled from a different function ``\texttt{sort}''. The training objective encourages the model to learn representations that bring the embeddings of $p$ and $a$ closer together while pushing the embeddings of $a$ and $n$ apart.}
\label{fig:ground-truth}
\end{figure}

\subsection{Ground Truth Construction via Compilation}
\label{sec:background:groundtruth}
As aforementioned, the core of ML-based BCSA methods is to learn high-quality binary code representations, which requires a large dataset of binary functions with known semantic relationships for training and evaluation.
To construct reliable ground truth for training and evaluation, existing works~\cite{xu2017neural,massarelli2019safe,ding2019asm2vec,wang2017imf,li2021palmtree,wang2022jtrans,wang2024clap,he2024hermessim} employ open-source software and compile the same source project under diverse configurations.
Given a project $P$ with source functions $\{s_1, s_2, \ldots, s_k\}$, the project is compiled with multiple configurations $C = \{c_1, c_2, \ldots, c_m\}$, where each configuration represents a unique combination of compiler (e.g., GCC, Clang), optimization level (e.g., \texttt{O0}, \texttt{O3}), and target architecture (e.g., x86\_64, AArch64).

Each configuration $c_i$ produces a set of binary functions (i.e., $\{b^{c_i}_1, b^{c_i}_2, \ldots, b^{c_i}_k\}$), where $b^{c_i}_j$ is compiled from source function $s_j$.
The debug symbol of $b^{c_i}_j$ serves as the identifier linking it back to $s_j$.
Two binary functions $b^{c_1}_i$ and $b^{c_2}_i$ sharing the same symbol (i.e., originating from the same source function $s_i$) form a \emph{positive pair}, while functions with different symbols form \emph{negative pairs}~\cite{wang2022jtrans,wang2024cebin,yu2020codecmr,ding2019asm2vec,luo2023vulhawk,wang2026vsim,li2022unleashing}.
\F~\ref{fig:ground-truth} shows that the positive pairs ($p$ and $a$) are built from the same source function ``\texttt{main}'', while the negative pair ($a$ and $n$) is formed by two functions compiled from different source functions (``\texttt{main}'' and ``\texttt{sort}'').
Apparently, the positive pairs share the same semantics, while negative pairs are likely dissimilar. Thus, the training objective is to learn representations that bring positive pairs closer together in the embedding space while pushing negative pairs apart.

Although the functions of positive pairs are compiled from the same source function, different configurations can lead to various compiler optimizations for produced assembly instructions, resulting in significant variations in the produced binary code, making the similarity analysis challenging~\cite{marcelli2022machine,rlobf2020}.
Thus, a careful design of the training objective is required to learn robust representations that capture the underlying semantics despite the syntactic differences.\looseness=-1

\subsection{Contrastive Learning for BCSA}
\label{sec:background:contrastive}

With positive and negative pairs established, models are often trained using contrastive learning to produce high-quality representations (i.e., embeddings).
Both loss functions described below measure the similarity between embeddings using cosine similarity~\cite{ding2019asm2vec,xu2017neural,wang2022jtrans,wang2024clap,yu2020codecmr}, i.e., $\text{cos}(\mathbf{u}, \mathbf{v}) = \frac{\mathbf{u} \cdot \mathbf{v}}{\|\mathbf{u}\| \|\mathbf{v}\|}$.

The embedding distance is typically defined using the cosine similarity, such as $d(\mathbf{u}, \mathbf{v}) = 1 - \text{cos}(\mathbf{u}, \mathbf{v})$.
The training objective encourages the model to produce embeddings where positive pairs have high cosine similarity (i.e., low distance) and negative pairs have low cosine similarity (i.e., high distance).

A commonly used loss function is the triplet loss~\cite{schroff2015facenet}:
\begin{equation}
    \mathcal{L}_{\text{triplet}} = \max\left(0, \; d(e_{a}, e_p) - d(e_{a}, e_n) + \alpha\right),
\end{equation}
where $a$ is an anchor, $p$ is a positive sample, $n$ is a randomly selected negative sample, and $\alpha$ is a margin hyperparameter.
The triplet loss penalizes cases where the anchor is closer to a negative sample than to the positive sample by at least a margin $\alpha$.

\begin{figure*}
\centering
\includegraphics[width=0.9\textwidth]{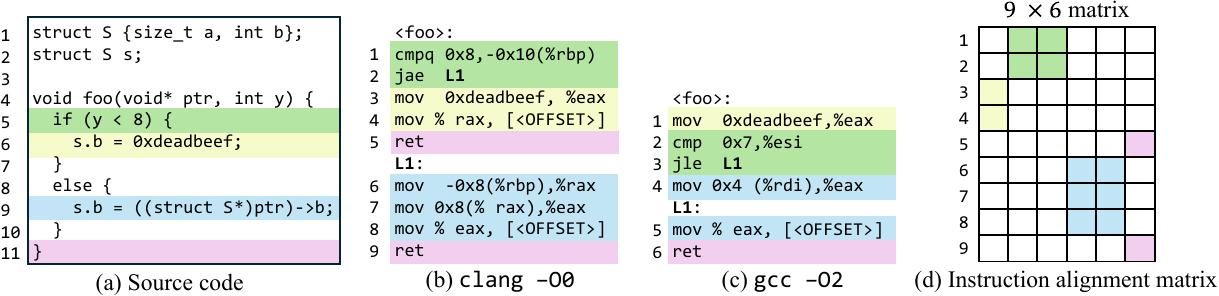}
\caption{Debug information provides a mapping from assembly instructions to source lines. (b) and (c) are compiled from the source code in (a) under different configurations. The colored instructions in (b) and (c) are originated from the source lines with the same color, thus they are semantically aligned. (d) is the instruction alignment matrix of (b) and (c), where the colored cells indicate the aligned instruction pairs with value of 1. Other white cells have value of 0.}
\label{fig:src-bin-mapping}
\end{figure*}

\subsection{Debug Information and Src-Bin Mapping}
\label{sec:background:debuginfo}

When compiling a program with debug flags (e.g., \texttt{-g}), the compiler generates debug information following standards such as DWARF~\cite{dwarf5}.
This debug information contains, among other data, a mapping from each assembly instruction to the source line that produced it.
\F~\ref{fig:src-bin-mapping} illustrates this mapping with a simple example.
The instructions of compiled binary functions in (b) and (c) can be mapped back to the source lines in (a) using the debug information.

Formally, for a binary function $B = [b_1, b_2, \ldots, b_n]$ compiled from source function $s$ with source lines $\{l_1, l_2, \ldots, l_m\}$, the debug information provides a mapping $\phi: b_i \mapsto l_j$, indicating that assembly instruction $b_i$ was generated from source line $l_j$.

Given two binary functions $B^{c_1}$ and $B^{c_2}$ compiled from the same source function $s$ under different configurations $c_1$ and $c_2$, instructions $b^{c_1}_i$ and $b^{c_2}_j$ are \emph{semantically aligned} if they map to the same source line, i.e., $\phi_{c_1}(b^{c_1}_i) = \phi_{c_2}(b^{c_2}_j)$.
This provides a fine-grained semantic correspondence at the instruction level, which is significantly richer than the coarse function-level matching used by existing methods.
 \section{Observation and Motivation}
\label{sec:observation}

In this section, we present a preliminary study of instruction alignment and the key observations that motivate our approach.
We conduct this study on \jtrans~\cite{wang2022jtrans}, a Transformer-based binary code embedding model that provides both a pre-trained checkpoint and a checkpoint finetuned for function-level BCSA using contrastive learning.
We evaluate both checkpoints on an instruction alignment task over function pairs compiled from the same source code under different compilation configurations.
\S\ref{sec:observation:embedding} and \S\ref{sec:observation:measurement} describe our instruction alignment evaluation methodology, and the results presented in \S\ref{sec:observation:comparison} motivate this study.

\subsection{Instruction Embedding}
\label{sec:observation:embedding}

\F~\ref{fig:overview} illustrates the instruction embedding process.
A binary code snippet is represented as a sequence of tokens fed into the transformer, which produces hidden states $\mathbf{H} = [\mathbf{h}_1, \mathbf{h}_2, \ldots, \mathbf{h}_S] \in \mathbb{R}^{S \times d}$, where $S$ is the sequence length and $d$ is the hidden dimension.
Since each assembly instruction typically spans multiple tokens (e.g., opcode and operands), we aggregate token-level representations into instruction-level embeddings.
We define an instruction index mapping $\psi: \{1, \ldots, S\} \to \{1, \ldots, K\} \cup \{-1\}$, where $K$ is the number of instructions in the function and $-1$ indicates special or padding tokens not belonging to any instruction.
The embedding of the $k$-th instruction is computed by mean-pooling the hidden states of all tokens belonging to that instruction:
\begin{equation}
    \mathbf{e}_k = \frac{1}{|\mathcal{T}_k|} \sum_{t \in \mathcal{T}_k} \mathbf{h}_t, \quad \text{where} \quad \mathcal{T}_k = \{t \mid \psi(t) = k\}.
\end{equation}
This process introduces no additional learnable parameters, and the instruction embeddings are derived directly from the transformer backbone.

Given two functions $A$ and $B$, we obtain their instruction embeddings $\{\mathbf{e}^A_1, \ldots, \mathbf{e}^A_n\}$ and $\{\mathbf{e}^B_1, \ldots, \mathbf{e}^B_m\}$, where $n$ and $m$ are the number of instructions in $A$ and $B$, respectively. The similarity between any instruction pair $(\mathbf{e}^A_i, \mathbf{e}^B_j)$ can be computed using cosine similarity, which serves as the basis for instruction alignment evaluation.

\subsection{Instruction Alignment Measurement}
\label{sec:observation:measurement}

Before leveraging instruction-level knowledge for training, we first investigate how well existing binary code embedding models capture instruction-level semantics.
Since binary functions are composed of assembly instructions, a natural question arises: given a pair of functions with identical symbols (i.e., compiled from the same source), can the model's instruction-level embeddings explain \emph{why} these two functions are considered similar?

We formulate instruction alignment as a retrieval task.
Given two binary functions $A = [a_1, a_2, \ldots, a_n]$ and $B = [b_1, b_2, \ldots, b_m]$ compiled from the same source function under different configurations, we establish an instruction alignment matrix $\mathbf{M} \in \{0, 1\}^{n \times m}$ using the debug information:
$\mathbf{M}_{i,j} = 1$ if $a_i$ and $b_j$ are semantically aligned (i.e., $\phi(a_i) = \phi(b_j)$), and $\mathbf{M}_{i,j} = 0$ otherwise. \F~\ref{fig:src-bin-mapping}(d) shows an example of such a matrix, where the colored cells indicate the aligned instruction pairs from \F~\ref{fig:src-bin-mapping}(b) and \ref{fig:src-bin-mapping}(c).

We first identify all instructions in $A$ that have at least one semantically aligned counterpart in $B$:
\begin{equation}
    \mathcal{I}^{A \to B} = \{a_i \mid \sum_{j=1}^{m} \mathbf{M}_{i,j} > 0 \}.
    \label{eq:insn-align-set}
\end{equation}
This filtering is necessary because compiler optimizations may eliminate certain instructions or inline callee semantics, leaving some instructions without a counterpart in the other binary function.
For each $a_i \in A'$, we rank all instructions in $B$ by the cosine similarity between their embeddings and $a_i$'s embedding, yielding a ranked list $B_{a_i}$.
We define the rank of $a_i$ as the highest position of any aligned instruction:
\begin{equation}
\text{rank}(B_{a_i}) = \min \left\{\text{pos}(b_j, B_{a_i}) \mid \mathbf{M}_{i,j} = 1\right\}
\end{equation}
where $\text{pos}(b_j, B_{a_i})$ denotes the position of $b_j$ in the ranked list $B_{a_i}$.
Intuitively, when an aligned instruction $b_j$ has a similar embedding to $a_i$, it ranks near the top, resulting in a small $\text{rank}(B_{a_i})$.

We evaluate the alignment quality using standard retrieval metrics, i.e., Mean Reciprocal Rank (MRR) and Recall@1:
\begin{equation}
    \text{Recall}^{\text{insn}}\text{@}1 = \frac{1}{|\mathcal{I}^{A \to B}|} \sum_{a_i \in \mathcal{I}^{A \to B}}\mathbb{1}\!\left(\text{rank}(B_{a_i}) = 1\right)
\end{equation}
\begin{equation}
    \text{MRR}^{\text{insn}} = \frac{1}{|\mathcal{I}^{A \to B}|} \sum_{a_i \in \mathcal{I}^{A \to B}}\frac{1}{\text{rank}(B_{a_i})}
\end{equation}
\noindent where $\mathbb{1}(\cdot)$ is the indicator function, which returns 1 if the condition holds and 0 otherwise.
Higher values of Recall@$1$ and MRR indicate better instruction alignment accuracy.

\begin{table}
\centering
\caption{Instruction alignment performance of pre-trained and finetuned \jtrans models on their test dataset.}
\label{tab:alignmentresults}
{\small
\begin{tabular}{lcc|cc}
\toprule
\multirow{2}{*}{\textbf{Model}} & \multicolumn{2}{c|}{\textbf{\mrrinsn}} & \multicolumn{2}{c}{\textbf{\rinsn}} \\
& \textbf{O0$\to$O2} & \textbf{O0$\to$O3} & \textbf{O0$\to$O2} & \textbf{O0$\to$O3} \\
\hline
Pre-trained & 0.635 & 0.630 & 0.524 & 0.519 \\
Finetuned   & 0.684 & 0.681 & 0.574 & 0.573 \\
\hline
& \textbf{O2$\to$O0} & \textbf{O3$\to$O0} & \textbf{O2$\to$O0} & \textbf{O3$\to$O0} \\
\hline
Pre-trained & 0.716 & 0.708 & 0.627 & 0.619 \\
Finetuned   & 0.731 & 0.722 & 0.641 & 0.632 \\
\hline
\end{tabular}}
\end{table}

\subsection{Preliminary Instruction Alignment Study}
\label{sec:observation:comparison}

\T~\ref{tab:alignmentresults} presents the instruction alignment performance of both the pre-trained and finetuned \jtrans models across different optimization level pairs.
The finetuned model consistently outperforms the pre-trained model in both MRR and Recall@1, indicating that the contrastive training at the function level has indeed improved the model's ability to capture instruction-level semantics.
The improvement in instruction alignment correlates with the finetuned model's superior performance on function-level BCSA, suggesting that better instruction-level understanding contributes to more accurate function-level similarity judgments.

Additionally, the results between non-optimized and lower optimized functions (i.e., O0$\to$O2 and O2$\to$O0) are better than the alignment between higher optimized settings (i.e., O0$\to$O3 and O3$\to$O0), which is expected since aggressive optimizations can largely alter the instruction sequence, making alignment more challenging.
This phenomenon further underscores the importance of instruction-level understanding for robust binary code similarity analysis, especially in scenarios involving heavily optimized binaries.\looseness=-1

\subsection{Motivation}
\label{sec:observation:motivation}

The above observations lead to our central research question: \emph{Can we explicitly incorporate instruction alignment knowledge into the training process to further improve binary code embeddings?}

If function-level training already implicitly improves instruction alignment, then directly optimizing for instruction alignment should provide an even stronger supervisory signal.
This fine-grained objective encourages the model to learn precise semantic correspondences at the instruction level, which should in turn produce higher-quality function-level embeddings.

Furthermore, explicit instruction alignment training may offer an additional benefit: \emph{interpretability}.
A model trained with instruction alignment can not only determine that two functions are similar with function-level embedding distance, but also pinpoint \emph{which specific instructions} correspond to each other, giving inspectable evidence alongside the similarity judgment.
 \section{Methodology}
\label{sec:design}

Previous \S\ref{sec:observation:embedding} has shown the instruction embedding mechanism, and this section explains the overview (\S\ref{sec:design:overview}) of \tool, the design of instruction alignment loss (\S\ref{sec:design:infonce}), the combined training objective (\S\ref{sec:design:training}), and the training data construction (\S\ref{sec:design:data}).\looseness=-1

\subsection{Overview}
\label{sec:design:overview}

\tool extends existing Transformer-based binary code embedding models, including \jtrans~\cite{wang2022jtrans} and \jclap~\cite{wang2024clap}, by augmenting standard function-level contrastive training with an instruction-level alignment objective derived from fine-grained source-line correspondences (\S\ref{sec:background:debuginfo}).
Although designed for Transformer-based models, the framework is also applicable to other architectures that can produce fine-grained embeddings.

\begin{figure}
\centering
\includegraphics[width=1.02\linewidth]{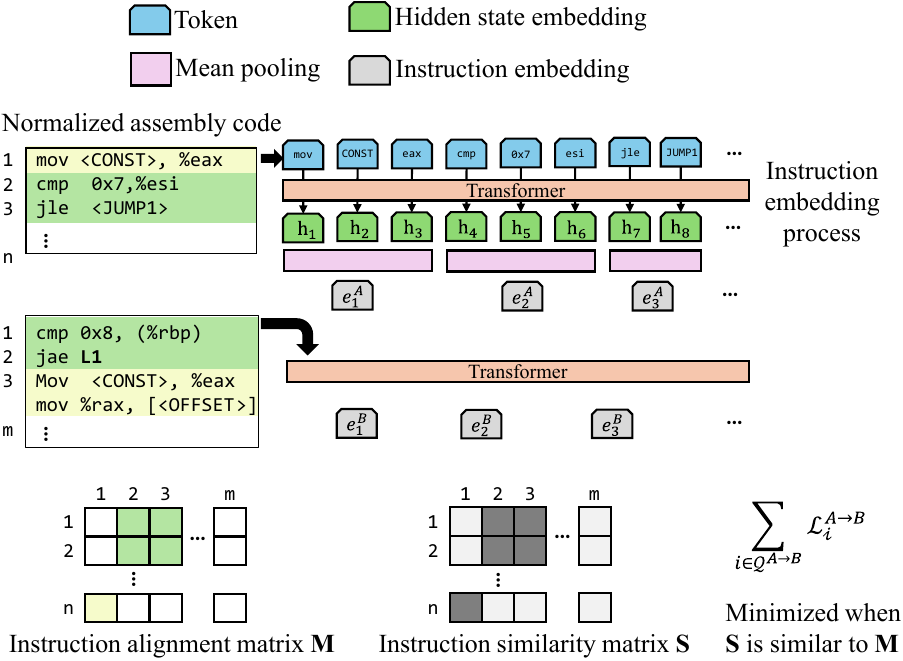}
\caption{Instruction alignment overview.
Colored $\mathbf{M}_{i,j}$ denotes instruction $a_i$ and $b_j$ originates from the same source line.
Deep gray $\mathbf{S}_{i,j}$ denotes $\text{cos}(\mathbf{e}^A_i, \mathbf{e}^B_j)$ is close to 1.}
\label{fig:overview}
\end{figure}

\F~\ref{fig:overview} illustrates the overall training framework.
Given a pair of binary functions $(A, B)$ compiled from the same or overlapping source code, our framework first constructs an instruction alignment matrix $\mathbf{M}$ using debug information.
It then encodes both functions with a shared Transformer to obtain token-level hidden states and function-level embeddings.
Initial instruction embeddings are derived by mean-pooling the hidden states of the tokens belonging to each instruction (\S\ref{sec:observation:embedding}).
Based on these instruction embeddings, an instruction similarity matrix $\mathbf{S}$ is computed using cosine similarity.
We then formulate an InfoNCE loss~\cite{oord2018representation} over $\mathbf{M}$ and $\mathbf{S}$ to optimize instruction alignment.
This loss is jointly combined with the function-level similarity objective, implemented as the triplet loss~\cite{schroff2015facenet}, to improve overall embedding quality.

\subsection{Instruction Alignment Loss}
\label{sec:design:infonce}

Given two binary functions $A$ and $B$ compiled from the same source function under different configurations, we obtain their instruction embeddings $\{\mathbf{e}^A_1, \ldots, \mathbf{e}^A_n\}$ and $\{\mathbf{e}^B_1, \ldots, \mathbf{e}^B_m\}$ through the instruction pooling module.
Using the debug information, we construct a binary match matrix $\mathbf{M} \in \{0, 1\}^{n \times m}$, where $\mathbf{M}_{ij} = 1$ if instructions $\mathbf{e}^A_i$ and $\mathbf{e}^B_j$ originate from the same source line.

\smallskip
{
\parh{Why InfoNCE?}
Unlike function-level BCSA, where negatives can be drawn from the entire dataset (e.g., millions of candidate functions), instruction alignment operates within a single function pair: for a query instruction $a_i$, the candidate set is simply the $m$ instructions of the paired function $B$.
Because a binary function contains only a bounded number of instructions, this candidate set is small enough to serve directly as the InfoNCE~\cite{oord2018representation} denominator, casting alignment as an $m$-way classification that pulls each query toward its matches and away from the remaining instructions without any negative sampling.
Moreover, instruction alignment is inherently many-to-many;
a single source line may expand into multiple instructions (e.g., through macro expansion), and multiple lines may share instructions.
Thus, we adopt a multi-positive generalization inspired by supervised contrastive learning~\cite{khosla2020supervised}: all instructions $b_j$ with $\mathbf{M}_{ij}=1$ are treated as positives, and the loss (\E~\ref{eq:infonce}) maximizes the aggregate probability mass over this positive set rather than forcing a single hard match.
}

\smallskip
\parh{Loss Formulation.}
We compute the instruction-to-instruction similarity matrix:
\begin{equation}
\mathbf{S}_{ij} = {\text{cos}(\mathbf{e}^A_i, \mathbf{e}^B_j)}.
    \label{eq:insn-cos}
\end{equation}

The alignment loss is computed symmetrically in both directions.
In the forward direction ($A \to B$), for each instruction $\mathbf{e}^A_i$ that has at least one match in $B$ (i.e., $\sum_j \mathbf{M}_{ij} > 0$), we compute:
\begin{equation}
    \mathcal{L}^{A \to B}_i = \log \sum_{j=1}^{m} \exp(\mathbf{S}_{ij}/\tau) - \log \sum_{j: \mathbf{M}_{ij}=1} \exp(\mathbf{S}_{ij}/\tau),
\label{eq:infonce}
\end{equation}
where $\tau$ is a temperature parameter.
Intuitively, \E~\ref{eq:infonce} is minimized when the model assigns all probability mass to the positive matches $\{j : \mathbf{M}_{ij} = 1\}$ among the $m$ candidates.
When there is exactly one positive match, Equation~\ref{eq:infonce} reduces to the standard InfoNCE loss~\cite{oord2018representation}; when multiple positives exist, it generalizes to a multi-positive variant that maximizes the aggregate softmax probability over the entire positive set.

The backward direction ($B \to A$) is computed analogously by transposing $\mathbf{S}$ and $\mathbf{M}$.
The total instruction alignment loss is the average over all valid queries in both directions:
\begin{equation}
    \mathcal{L}_{\text{align}} = \frac{1}{|\mathcal{I}^{A \to B}| + |\mathcal{I}^{B \to A}|} \left( \sum_{i \in \mathcal{I}^{A \to B}} \mathcal{L}^{A \to B}_i + \sum_{j \in \mathcal{I}^{B \to A}} \mathcal{L}^{B \to A}_j \right),
\end{equation}
where $\mathcal{I}^{A \to B}$ and $\mathcal{I}^{B \to A}$ are the sets of query instructions with at least one positive match defined by \E~\ref{eq:insn-align-set}.
The symmetric computation ensures that the alignment is bidirectional: instructions in $A$ are encouraged to find their counterparts in $B$, and vice versa.

\subsection{Combined Training Objective}
\label{sec:design:training}

\parh{Combined Loss.}
The overall training objective combines the whole function-level triplet loss with the instruction-level alignment loss:
\begin{equation}
    \mathcal{L} = \mathcal{L}_{\text{triplet}} + \lambda \cdot \mathcal{L}_{\text{align}},
\end{equation}
where $\lambda$ is a weighting coefficient that controls the relative importance of the alignment loss.

The function-level triplet loss $\mathcal{L}_{\text{triplet}}$ operates on the function-level embeddings (\S\ref{sec:background:contrastive}), encouraging semantically similar functions to have close embeddings while pushing dissimilar ones apart.
The instruction alignment loss $\mathcal{L}_{\text{align}}$ operates on the per-instruction embeddings, enforcing fine-grained semantic correspondences between individual instructions.

\smallskip
\parh{Layer Freezing.}
To reduce computational cost while preserving the pre-trained knowledge, we adopt a layer freezing strategy: the embedding layer and the first $L$ encoder layers of the BERT backbone are frozen during training, and only the upper layers are finetuned. This behavior is consistent with the finetuning stage of the original \jtrans model, which also freezes the lower layers during contrastive training~\cite{wang2022jtrans}.
The freezing strategy allows the model to retain the general binary code understanding acquired during pre-training while adapting the upper representations for instruction-level alignment.

\subsection{Training Data Construction}
\label{sec:design:data}

We construct training pairs from the compiled binaries with debug information, where each positive pair $(A, B)$ consists of two binary functions that share the same source function but are compiled under different configurations (e.g., different optimization levels or compilers).
Since both functions originate from the same source, their match matrix $\mathbf{M}$ is constructed by aligning all instructions that share the same source line (e.g., the $\mathbf{M}$ of \F~\ref{fig:overview}).
Positive pairs drive both the function-level triplet loss (as the positive example) and the instruction alignment loss (over all shared source lines).
A negative function is randomly sampled from a different symbol group to provide the negative example for the triplet loss.
Specifically, we reuse the binaries of BinaryCorp (i.e., the training data) of \jtrans, for fair comparison.
\label{sec:experiment}

\subsection{Baselines}
\label{sec:experiment:baselines}
To evaluate the effectiveness of instruction alignment, we compare our newly trained model directly against the models finetuned for BCSA of \jtrans~\cite{wang2022jtrans} and \jclap~\cite{wang2024clap}.

\smallskip
\parh{\jtrans} is a customized Transformer for learning jump-aware binary code embedding.
It outperforms prior models, including Genius~\cite{feng2016scalable}, Gemini~\cite{xu2017neural}, SAFE~\cite{massarelli2019safe}, Asm2vec~\cite{ding2019asm2vec}, OrderMatters~\cite{yu2020order}.
To perform a fair comparison, we continue training the finetuned \jtrans checkpoint with the auxiliary instruction alignment loss, using the same dataset, tokenization, and training settings as in the original \jtrans paper.\looseness=-1

\smallskip
\parh{\jclap} uses the RoBERTa base architecture~\cite{liu2019roberta}, a different architecture compared with \jtrans. It also employs a cross-modality training paradigm to align embeddings between binary code and natural language descriptions.
With the high-quality natural language embeddings, \jclap achieves the state-of-the-art performance on BCSA.\looseness=-1

\subsection{Datasets}
\label{sec:experiment:dataset}
To evaluate the effect of instruction alignment fairly, we use datasets adopted by prior BCSA studies, including BinaryCorp~\cite{wang2022jtrans} for training and BinKit~\cite{kim2022revisiting} for evaluation. All binaries are stripped before extracting assembly instructions.

\smallskip
\parh{BinaryCorp Dataset.}
To enable a fair comparison with \jtrans, we reuse its training dataset---the training partition of BinaryCorp dataset---for training.
However, not all binaries in BinaryCorp were compiled with debug information, which is required to construct ground truth for instruction alignment training.
Therefore, we use only the subset of BinaryCorp that contains debug information.
This training dataset contains 1,655,011 binary functions from 1,544 projects, forming 2,326,328 positive function pairs.

To evaluate the model trained with the auxiliary instruction alignment loss, we do not use the original BinaryCorp test set from the \jtrans paper, in order to avoid potential data leakage.
Specifically, the original test set contains binaries compiled from the same projects as those in the training set, so some test functions originate from the same source lines as training functions, which could inflate instruction alignment performance.
Instead, we construct the test set from BinKit~\cite{kim2022revisiting}.
We have manually checked the projects in BinKit and confirmed that none of them overlap with the projects in BinaryCorp, ensuring a clean evaluation of instruction alignment without data leakage.

\smallskip
\parh{BinKit Dataset.}
BinKit~\cite{kim2022revisiting} is a large-scale BCSA dataset containing 213,400 binaries compiled from 51 open-source projects under diverse compiler versions and compilation settings.
All binaries in BinKit include debug information, making the dataset suitable for evaluating instruction alignment.
However, as noted in prior work~\cite{wang2026vsim,wang2022jtrans}, BinKit suffers from substantial duplication, where many binary functions are compiled from the same source lines.
For example, the Coreutils project accounts for 42.1\% of all binaries in the dataset, while it merely contributes nearly 2,700 source functions.
To mitigate this issue, we follow the deduplication preprocess of \textsc{vSim}~\cite{wang2026vsim}.
Specifically, for each compilation configuration, functions from the same project that share the same symbol are treated as duplicates; one function is retained for evaluation, and the remaining duplicates are removed.
We also remove 18 projects (e.g., gawk) that exist in the training set of BinaryCorp.
After the preprocess, we use the binary functions of Coreutils project for validation, and the binary functions of the remaining 32 projects as the test set. We use the binaries produced by the latest and oldest versions of GCC and Clang (gcc-11, clang-13, gcc-4.9, clang-4) to cover a wide range of compiler behaviors and avoid duplications.

\subsection{Metrics}
\label{sec:experiment:metrics}

For direct comparison with prior work on binary code similarity analysis (BCSA), we adopt the widely used Recall@1~\cite{marcelli2022machine,wang2022jtrans,ding2019asm2vec,he2024hermessim,wang2024clap}, which measures the proportion of queries whose true match is ranked first.

Let the query set be $\mathcal{Q} = \{q_1, q_2, \dots, q_N\}$, and let $\text{rank}(q_i)$ denote the rank of the ground-truth match for query function $q_i$ in the candidate pool $\mathcal{P}$.
Then, \rfunc is defined as
\begin{equation}
    \text{Recall}^{\text{func}}@1 = \frac{1}{N} \sum_{f \in \mathcal{Q}} \mathbb{1}(rank(f) = 1),
\end{equation}
\noindent where $\mathbb{1}(\cdot)$ is the indicator function.
A higher \rfunc indicates better retrieval accuracy and higher-quality function-level embeddings.
Previous studies~\cite{ding2019asm2vec,wang2022jtrans,wang2024clap,wang2026vsim,wang2024cebin} also report mean reciprocal rank (MRR);
we omit it for simplicity, because MRR is strongly positively correlated with \rfunc in the BCSA.\looseness=-1

\subsection{Implementation Details}
\label{sec:experiment:implementation}

All experiments are conducted on a server with an NVIDIA A6000 GPU (48\,GB), 256\,GB RAM, and an AMD Threadripper 3970X CPU.
For fair comparison, we continue training the finetuned \jtrans and \jclap checkpoints with the auxiliary instruction alignment loss, using the same disassembler (i.e., IDA Pro~\cite{ida}), the training dataset (\S\ref{sec:experiment:dataset}), and training settings as in the original \jtrans paper~\cite{wang2022jtrans}.

We set the learning rate to $1e-5$, the frozen layer count $L$ to 10, the weight $\lambda$ of the instruction alignment loss to $0.001$, and train the models two epochs.
The former two hyperparameters are consistent with the original \jtrans finetuning settings, while the last one is chosen empirically based on preliminary experiments to balance the two objectives effectively.
We choose a small $\lambda$ because the finetuned \jtrans checkpoint already provides strong function-level similarity performance.
In practice, the original triplet loss is extremely small (below 0.001), whereas the instruction alignment loss is much larger (around 1.5).
A small $\lambda$ is therefore necessary to balance the two objectives and prevent instruction-level alignment from overwhelming function-level representation learning~\cite{chen2018gradnorm}. \section{Evaluation}
\label{sec:evaluation}

To measure the effectiveness of instruction alignment training, we conduct a comprehensive evaluation addressing the following research questions (RQs):
\begin{packeditemize}
    \item \textbf{RQ1}: How does instruction alignment training converge, and how does it affect instruction-level alignment quality?
    \item \textbf{RQ2}: How does instruction alignment training impact function-level BCSA performance?
    \item \textbf{RQ3}: Does the instruction-level alignment signal provide stronger discriminability than function-level embeddings?
\item {\textbf{RQ4}: How reliable are compiler-generated labels as a training signal, and how resilient is our method to label noise?}
    \item {\textbf{RQ5}: How effective is the auxiliary objective when combined with hard negative mining?}
\end{packeditemize}
Moreover, we present case studies (\S\ref{sec:evaluation:casestudy}) to understand \tool's difficult situations and show the potential for patch presence detection, a challenging application for function-level representations.

\subsection{RQ1: Training Loss Convergence}
\label{sec:evaluation:rq1}

\begin{figure}[t]
    \centering
    \includegraphics[width=\linewidth]{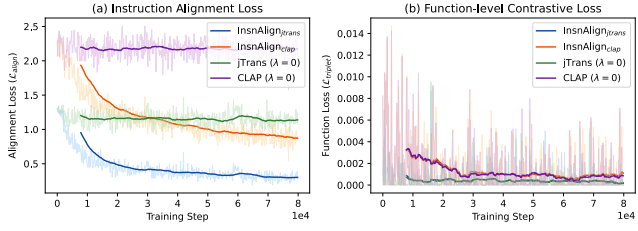}
    \caption{Training loss curves. (a) Instruction alignment loss $\mathcal{L}_{align}$ decrease when $\lambda > 0$. (b) Function-level contrastive loss $\mathcal{L}_{triplet}$ remain near zero, indicating negligible degradation.
    Note that the instruction alignment loss is much greater than the function-level loss.}
    \label{fig:training-loss}
\end{figure}

\F~\ref{fig:training-loss} shows the training loss curves over the course of two epochs (approximately 80K training steps).
We observe several key trends:

\smallskip
\parh{Alignment Loss Decreases Significantly.}
When $\lambda > 0$, the instruction alignment loss $\mathcal{L}_{align}$ drops sharply during the first epoch (\F~\ref{fig:training-loss}(a)).
The rapid initial decline indicates that the models quickly learn to align instruction-level embeddings across function pairs compiled from the same source.
The loss continues to decrease in the second epoch but at a slower rate, suggesting convergence of the alignment objective using those pre-trained models.
Moreover, we also finetune \jtrans and \jclap with $\lambda = 0$ for isolating the effect of instruction alignment training.
As shown in \F~\ref{fig:training-loss}, their $\mathcal{L}_{align}$ remains high.\looseness=-1

\smallskip
\parh{Function-level Loss Remains Stable.}
The function-level contrastive loss $\mathcal{L}_{func}$ (\F~\ref{fig:training-loss}(b)) remains near zero throughout training, consistent with the fact that the models are initialized from well-trained checkpoints that have already converged on the function-level objective.
The loss values of \toolj and \jtrans ($\lambda = 0$) are stable, and 
the loss values of \toolc and \jclap ($\lambda = 0$) decrease slightly.
This stability confirms that the auxiliary instruction alignment loss does not degrade the model's existing function-level similarity capability.
The following evaluation uses the models further trained with $\lambda = 0$ as baselines, since they slightly outperform their original versions.\looseness=-1

\begin{figure}
\centering
\includegraphics[width=1.0\linewidth]{figures/insn\_align\_bar\_v2\_x1.pdf}
\caption{{Average \rinsn and \mrrinsn.}}
\label{fig:rq2}
\end{figure}

\smallskip
\parh{Instruction Alignment Results.}
We evaluate the instruction alignment quality using the \mrrinsn and \rinsn metrics defined in \S\ref{sec:observation:measurement}. \F~\ref{fig:rq2} presents the results between binaries compiled with trivial optimization and aggressive optimization of BinKit dataset.
The models trained with our auxiliary task significantly outperform those without by 50.9\% for \jtrans and 88.2\% for \jclap on \rinsn, demonstrating that our training method can significantly improve the instruction alignment performance. 
\subsection{RQ2: Impact on Binary Function Embedding}
\label{sec:evaluation:rq2}

This RQ investigates the impact of instruction alignment training on function-level representation learning. We measure the \rfunc on the BinKit~\cite{kim2022revisiting} dataset across 16 cross-compiler, cross-optimization-level settings (4 compilers at O0 vs. O3).\looseness=-1

\subsubsection{Setup}

We compare four model variants: the \jtrans and \jclap models trained with $\lambda = 0$, and the models being trained with instruction alignment task (\tool uses $\lambda > 0$).
We evaluate on a pool size of 10,000 functions per comparison setting, consistent with the most challenging setting in prior work~\cite{wang2022jtrans,wang2024clap,wang2026vsim}.

\subsubsection{Results}

\T~\ref{tab:rq3-baseline} presents the \rfunc for each model variant across all 16 comparison settings.

\begin{table}[t]
\centering
\caption{Function-level \rfunc across 16 cross-compiler, cross-optimization settings (O0 vs.\ O3).}
\label{tab:rq3-baseline}
\resizebox{\linewidth}{!}{
\begin{threeparttable}
\begin{tabular}{llcccc}
\toprule
O0 & O3 & jTrans ($\lambda=0$) & \toolj & CLAP ($\lambda=0$) & \toolc \\
\midrule
\multirow{4}{*}{GCC-11} & GCC-11 & 0.5011 & \textbf{0.5269} & 0.6507 & \textbf{0.6675} \\
 & Clang-13 & 0.4525 & \textbf{0.4729} & 0.6490 & \textbf{0.6702} \\
 & GCC-4 & 0.4768 & \textbf{0.5082} & 0.6296 & \textbf{0.6472} \\
 & Clang-4 & 0.4386 & \textbf{0.4630} & 0.6563 & \textbf{0.6745} \\
\midrule
\multirow{4}{*}{Clang-13} & GCC-11 & 0.3768 & \textbf{0.4088} & 0.6371 & \textbf{0.6548} \\
 & Clang-13 & 0.3687 & \textbf{0.3966} & 0.6467 & \textbf{0.6661} \\
 & GCC-4 & 0.3543 & \textbf{0.3886} & 0.6223 & \textbf{0.6450} \\
 & Clang-4 & 0.3526 & \textbf{0.3879} & 0.6523 & \textbf{0.6717} \\
\midrule
\multirow{4}{*}{GCC-4} & GCC-11 & 0.4803 & \textbf{0.5089} & 0.6324 & \textbf{0.6539} \\
 & Clang-13 & 0.4395 & \textbf{0.4509} & 0.6380 & \textbf{0.6587} \\
 & GCC-4 & 0.4909 & \textbf{0.5159} & 0.6474 & \textbf{0.6623} \\
 & Clang-4 & 0.4274 & \textbf{0.4451} & 0.6395 & \textbf{0.6585} \\
\midrule
\multirow{4}{*}{Clang-4} & GCC-11 & 0.3233 & \textbf{0.3542} & 0.6288 & \textbf{0.6480} \\
 & Clang-13 & 0.3180 & \textbf{0.3442} & 0.6389 & \textbf{0.6593} \\
 & GCC-4 & 0.3023 & \textbf{0.3375} & 0.6175 & \textbf{0.6384} \\
 & Clang-4 & 0.3142 & \textbf{0.3409} & 0.6470 & \textbf{0.6673} \\
\midrule
\multicolumn{2}{l}{Average} & 0.4011 & \textbf{0.4282} & 0.6396 & \textbf{0.6590} \\
\bottomrule
\end{tabular}
\end{threeparttable}
}
\end{table}

\smallskip
\parh{Improvements on Function Embeddings.}
Compared with the models trained with $\lambda = 0$, instruction alignment training can further improve the function-level embeddings, achieving higher \rfunc across all settings.
These results demonstrate that the auxiliary instruction alignment objective not only improves instruction-level representation quality (RQ1) but also benefits function-level representation learning, despite the function-level contrastive loss remaining stable during training.

\smallskip
\parh{Consistent Improvement Across Compiler Pairs.}
The improvement holds consistently across all 16 compiler configuration pairs, including both recent compilers (e.g., \texttt{gcc-11} and \texttt{clang-13}) and outdated compilers (e.g., \texttt{gcc-4.9} and \texttt{clang-4}).
As shown in \T~\ref{tab:rq3-baseline}, \jtrans performs much better on GCC-11 compilers, indicating the training data used in this study was compiled with GCC~\cite{wang2026vsim}, while the improvement on the comparison with unseen compilers (Clang-4 and Clang-13) is still significant, and \toolc\ shows even increases over \jclap across different settings.
This experiment demonstrates our approach is resilient to diverse compiler behaviors and optimization levels, denoting its robustness to distribution shift.\looseness=-1

\subsection{RQ3: Discriminability of Similar Functions}
\label{sec:evaluation:rq3}

Beyond retrieval accuracy, instruction alignment offers inspectable evidence:
because similarity is examined at the instruction level, the aligned instruction pairs constitute concrete, inspectable evidence for why two functions are judged similar or dissimilar.
{
For such evidence to be trustworthy, the instruction-level signal must itself be discriminative.
In this RQ, we therefore quantify to what extent it separates positive function pairs (compiled from the same source) from negative pairs (compiled from different sources).
}
This evaluation is distinct from the instruction retrieval task in RQ1: there, \mrrinsn and \rinsn measure the instruction embedding quality, whereas here we ask whether finer-grained similarities provide a more discriminative signal than the function-level embedding.\looseness=-1

We define a \emph{per-pair instruction alignment score} to aggregate instruction-level similarities into a single scalar that characterizes the alignment quality of a function pair.
{
We hypothesize that this finer-grained score can offer more discriminative and interpretable evidence for similarity judgments than the coarse-grained function embedding similarity, and test this hypothesis below.
}

\smallskip
\parh{Mean Alignment Score (MAS).}
Given two binary functions $A = [a_1, \ldots, a_n]$ and $B = [b_1, \ldots, b_m]$ with instruction similarity matrix $\mathbf{S}$ defined in \E~\ref{eq:insn-cos}, we compute the mean alignment score as:
\begin{equation}
    \text{MAS}(A, B) = \frac{1}{2} \left( \frac{1}{n} \sum_{i=1}^{n} \max_{1 \leq j \leq m} \mathbf{S}_{i,j} + \frac{1}{m} \sum_{j=1}^{m} \max_{1 \leq i \leq n} \mathbf{S}_{i,j} \right).
\end{equation}
The MAS captures the average best-match similarity between instructions of the two functions in both directions, providing a symmetric measure of how well the instructions align overall.
For positive pairs, we expect MAS to be high because many instructions originate from the same source lines and should have similar embeddings.
For negative pairs, instructions lack true semantic correspondences, so the best-match similarities should be lower on average.\looseness=-1

Similar to the cosine similarity between function-level embeddings, MAS itself is not a binary indicator of similarity; rather, it provides a continuous measure of how well the instructions align, which can serve as interpretable evidence for a similarity judgment. Additionally, we can measure the discriminability of both function embedding cosine similarity and MAS in distinguishing positive vs. negative pairs using the following metrics.

\subsubsection{Metrics}

We treat the distinction between positive and negative pairs as a binary classification problem and report two metrics.
\emph{AUC-ROC}~\cite{bradley1997use} measures how well MAS (or cosine similarity) separates positives from negatives: a value close to 1.0 indicates near-perfect discrimination, while 0.5 indicates no discriminability~\cite{saito2015precision}.
\emph{Cohen's $d$}~\cite{cohen2013statistical} quantifies the effect size between the score distributions of positive and negative pairs:
\begin{equation}
    d = \frac{\mu_{+} - \mu_{-}}{\sqrt{(\sigma_{+}^2 + \sigma_{-}^2) / 2}},
\end{equation}
where $\mu_{+}, \sigma_{+}$ and $\mu_{-}, \sigma_{-}$ are the mean and standard deviation for positive and negative pairs, respectively. A larger Cohen's $d$ indicates clearer separation.\looseness=-1

\subsubsection{Setup}

For each compilation configuration pair (e.g., O0$\to$O3), we construct positive pairs from functions sharing the same source symbol and sample an equal number of negative pairs.
We evaluate AUC-ROC and Cohen's $d$ for both MAS and function embedding cosine similarity ($cos$).

\smallskip
\parh{Negative Sampling Strategy.}
In real-world BCSA, positive and negative pairs are inherently imbalanced: each query has limited true match but potentially thousands of negatives. Random negative sampling therefore produces mostly trivially dissimilar pairs, inflating AUC for all models (\T~\ref{tab:auc})~\cite{bradley1997use,hand2023notes,saito2015precision}.
We adopt a harder strategy: for each query, negatives are drawn from the top-5 candidates by function-level cosine similarity that do not share the same symbol, stress testing discriminability against hard negatives that are close in the embedding space but semantically different.\looseness=-1

\begin{table}
\centering
\begin{threeparttable}
\caption{AUC and Cohen's $d$.}
\label{tab:auc}
{\small
\begin{tabular}{lccc|cc}
\toprule
\multirow{2}{*}{\textbf{Model}} & \textbf{Negative} & \multicolumn{2}{c|}{$\textbf{AUC}$} & \multicolumn{2}{c}{$\textbf{Cohen's }d$} \\
& \textbf{sampling} & \textbf{MAS} & \textbf{$cos$} & \textbf{MAS} & \textbf{$cos$} \\
\hline
\jtrans ($\lambda = 0$) & Top-5 & 0.644 & 0.573 & 0.435 & 0.111 \\
\toolj & Top-5 & 0.720 & 0.584 & 0.778 & 0.144 \\
\jclap ($\lambda = 0$) & Top-5 & 0.646 & 0.650 & 0.236 & 0.190 \\
\toolc & Top-5 & 0.702 & 0.698 & 0.622 & 0.526 \\
\hline
\jtrans ($\lambda = 0$) & Random & 0.986 & 0.996 & 3.606 & 4.487 \\
\toolj & Random & 0.974 & 0.996 & 2.974 & 4.621 \\
\jclap ($\lambda = 0$) & Random & 0.986 & 0.987 & 4.373 & 4.393 \\
\toolc & Random & 0.991 & 0.995 & 4.250 & 4.998 \\
\hline
\end{tabular}}
\begin{tablenotes}
\footnotesize
\item[1] The reported AUC and Cohen's $d$ are the average over seeds 3, 5, 7, 42.
\end{tablenotes}
\end{threeparttable}
\end{table}

\subsubsection{Results and Analysis}

\T~\ref{tab:auc} illustrates the AUC and Cohen's $d$ for both MAS and function-level cosine similarity ($cos$) under the different negative sampling strategies.

\smallskip
\parh{Instruction Alignment (MAS) vs. Function Embedding ($cos$).}
Under random negative sampling, both MAS and cosine similarity achieve high AUC for all models and are comparable ($cos$ is even marginally higher), as trivially dissimilar negatives are easy to reject at either granularity.
The distinction emerges only under the harder top-5 sampling, where both metrics drop: here MAS attains a higher Cohen's $d$ than $cos$ for every model and a higher AUC in most cases, indicating that the instruction-level signal is more discriminative when function embeddings are close.

\smallskip
\parh{Effect of Instruction Alignment Training.}
Comparing the models trained with $\lambda = 0$ and the models trained with \tool, we observe that instruction alignment training improves the AUC and Cohen's $d$ for both MAS and cosine similarity in the challenging setting. This observation is consistent with the evaluation of function-level BCSA performance in RQ2, where instruction alignment training also improves retrieval accuracy.
The improvement in AUC and Cohen's $d$ indicates that instruction alignment training not only enhances the quality of instruction embeddings but also yields clearer separation between similar and dissimilar pairs, making the instruction-level evidence a sharper and more reliable signal for explaining the model's similarity judgments.\looseness=-1

\subsubsection{A Synergy Effect}
\label{sec:rq4:synergy}
Because MAS shows stronger discriminability in distinguishing similar from dissimilar pairs than function-level embeddings, especially when two functions share close embeddings, we investigate whether combining the two signals can further improve function-level retrieval performance.
We define the \emph{synergy score} as a weighted combination:
\begin{equation}
    \text{Score}(A, B) = (1 - \gamma) \cdot \cos(\mathbf{f}_A, \mathbf{f}_B) + \gamma \cdot \text{MAS}(A, B),
\end{equation}
where $\mathbf{f}_A, \mathbf{f}_B$ are function embeddings, and $\gamma = 0.5$ balances the two signals.
Note that instruction alignment is time-consuming to compute, so we only apply the synergy scoring to the challenging comparison. We first rank candidates based on function-level cosine similarity, then compute MAS for the top 100 candidates and re-rank them using the synergy score. This approach is consistent with previous re-ranking approaches~\cite{wang2024cebin,wang2022enhancing}.
\T~\ref{tab:rq4-synergy} presents the synergy results.\looseness=-1

\begin{table}[t]
\centering
\caption{\rfunc with synergy scoring.}
\label{tab:rq4-synergy}
\resizebox{\linewidth}{!}{
\begin{threeparttable}
\begin{tabular}{llcccc}
\toprule
O0 & O3 & jTrans ($\lambda=0$) & \toolj & CLAP ($\lambda=0$) & \toolc \\
\midrule
\multirow{4}{*}{GCC-11} & GCC-11 & 0.5561 & \textbf{0.5996} & 0.6577 & \textbf{0.6847} \\
 & Clang-13 & 0.5147 & \textbf{0.5757} & 0.6574 & \textbf{0.6857} \\
 & GCC-4 & 0.5295 & \textbf{0.5772} & 0.6364 & \textbf{0.6607} \\
 & Clang-4 & 0.5085 & \textbf{0.5715} & 0.6643 & \textbf{0.6893} \\
\midrule
\multirow{4}{*}{Clang-13} & GCC-11 & 0.4599 & \textbf{0.5448} & 0.6429 & \textbf{0.6715} \\
 & Clang-13 & 0.4574 & \textbf{0.5373} & 0.6537 & \textbf{0.6778} \\
 & GCC-4 & 0.4445 & \textbf{0.5236} & 0.6305 & \textbf{0.6569} \\
 & Clang-4 & 0.4479 & \textbf{0.5352} & 0.6618 & \textbf{0.6850} \\
\midrule
\multirow{4}{*}{GCC-4} & GCC-11 & 0.5370 & \textbf{0.5800} & 0.6411 & \textbf{0.6698} \\
 & Clang-13 & 0.5022 & \textbf{0.5606} & 0.6425 & \textbf{0.6733} \\
 & GCC-4 & 0.5418 & \textbf{0.5860} & 0.6547 & \textbf{0.6746} \\
 & Clang-4 & 0.4954 & \textbf{0.5571} & 0.6492 & \textbf{0.6733} \\
\midrule
\multirow{4}{*}{Clang-4} & GCC-11 & 0.4097 & \textbf{0.5106} & 0.6352 & \textbf{0.6645} \\
 & Clang-13 & 0.4131 & \textbf{0.5040} & 0.6484 & \textbf{0.6749} \\
 & GCC-4 & 0.3931 & \textbf{0.4915} & 0.6233 & \textbf{0.6520} \\
 & Clang-4 & 0.4102 & \textbf{0.5059} & 0.6522 & \textbf{0.6795} \\
\midrule
\multicolumn{2}{l}{Average} & 0.4763 & \textbf{0.5475} & 0.6470 & \textbf{0.6733} \\
\multicolumn{2}{l}{Average improvement\footnotemark[1]} & +18.7\% & \textbf{+27.9\%} & +1.16\% & \textbf{+2.17\%} \\
\bottomrule
\end{tabular}
\begin{tablenotes}
    \item[1] The average improvements over the values shown in \T~\ref{tab:rq3-baseline}.
\end{tablenotes}
\end{threeparttable}
}
\end{table}

\smallskip
\parh{Synergy Consistently Improves Retrieval.}
For \toolj, synergy scoring increases \rfunc from 0.4282 to 0.5475, yielding a 27.9\% improvement, whereas \toolc gains only 2.17\%.
This discrepancy is consistent with the difference of MAS and $cos$ in discriminability shown in \T~\ref{tab:auc}. With the negative sampling strategy, MAS and $cos$ of \toolc are similar, while \toolj's MAS is significantly larger than $cos$.
The original \jtrans and \jclap models also benefit from synergy scoring, indicating that instruction-level alignment provides complementary information regardless of the training stage.
Moreover, our instruction alignment training further strengthens this benefit.
As shown in \T~\ref{tab:rq4-synergy}, the relative improvement increases from 18.7\% to 27.9\% for the \jtrans-based model and from 1.16\% to 2.17\% for the \jclap-based model, even though the trained models already achieve higher base \rfunc.

\subsection{RQ4: Label Quality and Noise Resilience}
\label{sec:evaluation:rq4}
{
\tool relies on compiler-generated debug information to establish instruction-level correspondences for training.
Although debug information provides valuable fine-grained supervision, it is not perfectly precise: aggressive compiler optimizations may attribute instructions to coarse or imprecise source lines~\cite{lu2024dtd,unseendelta2026issta}.
This RQ asks whether such silver labels remain useful when they contain noise.
To assess the quality of the compiler-generated labels, we analyze two aspects: coverage and correctness.

\smallskip
\parh{Coverage.}
Because \texttt{-O0} binaries undergo minimal optimization, we use them as the reference for measuring coverage in optimized binaries.
We consider an \texttt{-O0} instruction covered if it has at least one corresponding instruction in the optimized binary.
Across our training data, {86.6\%} of instructions in \texttt{-O0} functions are covered in their \texttt{-O3} counterparts.
This result suggests that the compiler-generated labels retain broad coverage even under aggressive optimization.
}\looseness=-1

\begin{figure}
    \centering
    \begin{minipage}{0.6\linewidth}
\begin{lstlisting}[numbers=left]
else if(!(in=fopen(file_name,"r")))
  return -1;
if(!use_stdin && fclose(in)!=0)
  rc = -1;
return rc;
\end{lstlisting}
    \end{minipage}
    \hfil
    \begin{minipage}{0.35\linewidth}
\begin{lstlisting}[numbers=left]
  mov eax, 0FFFFFFFFh
  jmp returnLabel
returnLabel:
  ret
\end{lstlisting}
    \end{minipage}
\caption{Example of a \textit{PLAUSIBLE} label.
Lines 2, 4, and 5 are absorbed into the common return sequence on the right; thus, the \texttt{mov} instruction can be attributed to both lines 2 and 4,
although the assembly is labeled only with line 2.}
\label{fig:plausible-example}
\end{figure}

\smallskip
{
\parh{Correctness.}
We assess correctness on \texttt{-O3} binaries, where aggressive optimization makes label misattribution most likely.
Given recent evidence that LLMs can analyze binary code~\cite{ICLR2025ef283d62,decllm2025,degpt2024,10795058}, we use an LLM to check all validation-set mappings and manually audit 100 randomly sampled functions (24,004 mappings across 4,470 compiled source lines).
The manual and LLM assessments are consistent.\looseness=-1

The mappings are classified as \textit{CORRECT} (86.8\%), \textit{PLAUSIBLE} (5.3\%), \textit{UNVERIFIABLE} (5.1\%), or \textit{SUSPICIOUS or WRONG} (2.8\%).
\textit{PLAUSIBLE} mappings are likely or partially correct but cannot be confirmed from assembly alone (\F~\ref{fig:plausible-example}); \textit{UNVERIFIABLE} mappings arise from imported glibc/gnulib headers or pure prologue/epilogue code.
With only 2.8\% classified as \textit{SUSPICIOUS or WRONG}, the compiler-generated labels appear highly reliable.
}\looseness=-1

\smallskip
{
\parh{Noise Resilience.}
We use \toolj, the strongest instruction-alignment variant in \S\ref{sec:evaluation:rq1}, and corrupt 5\%, 10\%, and 20\% of the instruction-to-source correspondences by randomly reassigning instructions to different source lines of the same function, then retrain and re-evaluate the model.
As shown in \F~\ref{fig:rq2}, label noise degrades alignment quality only moderately: even with 20\% injected noise, \toolj achieves 0.634 \rinsn, still 26.0\% higher than the baseline.
Thus, the auxiliary objective does not require perfectly clean debug labels.
Note that this random corruption is a stress-test proxy for noisy supervision; systematic mis-attribution caused by aggressive optimizations is separately bounded by the correctness analysis above.
}
 
\subsection{RQ5: Effectiveness with Hard Negatives}
\label{sec:evaluation:rq5}
{
Our main evaluation (RQ1-4) uses random negative sampling for the function-level triplet loss.
This design keeps the training protocol consistent across all settings and isolates the contribution of instruction alignment training.
This RQ further examines whether the auxiliary objective remains effective when the function-level contrastive objective is strengthened with hard negative mining~\cite{robinson2021contrastive}.
We use \jclap-based model since it shows better performance in our function retrieval experiments.

\smallskip
\parh{Setup.}
To mine hard negatives, we first encode all functions in the training set with the current model and retrieve the top-5 functions for each anchor according to cosine similarity.
During training, the negative sample in each triplet is drawn uniformly from this top-5 candidate set (true matches are excluded), and we refresh the mined top-5 candidates every 10,000 training steps.
Hard negatives make $\mathcal{L}_{triplet}$ substantially larger than in random negative sampling, so we increase $\lambda$ to $0.02$ to keep $\mathcal{L}_{triplet}$ and $\mathcal{L}_{align}$ balanced.}\looseness=-1

\begin{table}[t]
\centering
\caption{{\jclap-based models under hard-negative training.}}
\label{tab:rq5}
\resizebox{\linewidth}{!}{\begin{tabular}{llcccc}
\toprule
\multirow{2}{*}{O0} & \multirow{2}{*}{O3} & \multicolumn{2}{c}{\textbf{Non-synergy}} & \multicolumn{2}{c}{\textbf{Synergy}} \\
\cmidrule(lr){3-4} \cmidrule(lr){5-6}
 & & CLAP ($\lambda=0$) & \toolc & CLAP ($\lambda=0$) & \toolc \\
\midrule
\multirow{4}{*}{GCC-11} & GCC-11 & 0.6791 & \textbf{0.6939} & 0.6937 & \textbf{0.7075} \\
 & Clang-13 & 0.6827 & \textbf{0.6962} & 0.6982 & \textbf{0.7085} \\
 & GCC-4 & 0.6604 & \textbf{0.6698} & 0.6753 & \textbf{0.6841} \\
 & Clang-4 & 0.6882 & \textbf{0.7004} & 0.7000 & \textbf{0.7141} \\
\midrule
\multirow{4}{*}{Clang-13} & GCC-11 & 0.6690 & \textbf{0.6783} & 0.6853 & \textbf{0.6947} \\
 & Clang-13 & 0.6790 & \textbf{0.6876} & 0.6930 & \textbf{0.7039} \\
 & GCC-4 & 0.6582 & \textbf{0.6652} & 0.6720 & \textbf{0.6810} \\
 & Clang-4 & 0.6868 & \textbf{0.6956} & 0.7006 & \textbf{0.7091} \\
\midrule
\multirow{4}{*}{GCC-4} & GCC-11 & 0.6697 & \textbf{0.6777} & 0.6822 & \textbf{0.6930} \\
 & Clang-13 & 0.6714 & \textbf{0.6844} & 0.6855 & \textbf{0.6979} \\
 & GCC-4 & 0.6739 & \textbf{0.6837} & 0.6857 & \textbf{0.6990} \\
 & Clang-4 & 0.6755 & \textbf{0.6867} & 0.6903 & \textbf{0.6998} \\
\midrule
\multirow{4}{*}{Clang-4} & GCC-11 & 0.6647 & \textbf{0.6735} & 0.6790 & \textbf{0.6891} \\
 & Clang-13 & 0.6711 & \textbf{0.6839} & 0.6874 & \textbf{0.7004} \\
 & GCC-4 & 0.6516 & \textbf{0.6593} & 0.6673 & \textbf{0.6777} \\
 & Clang-4 & 0.6809 & \textbf{0.6907} & 0.6947 & \textbf{0.7051} \\
\midrule
\multicolumn{2}{l}{Average} & 0.6726 & \textbf{0.6829} & 0.6869 & \textbf{0.6978} \\
\bottomrule
\end{tabular}}
\end{table}

\smallskip
{
\parh{Results.}
As shown by \T~\ref{tab:rq5}, with hard negative mining enabled, function-level retrieval performance improves, compared to the random negative sampling results (\T~\ref{tab:rq3-baseline}).
Notably, \tool continues to improve function-level retrieval over the hard-negative baseline, and synergy re-ranking (\S\ref{sec:rq4:synergy}) provides a further gain.
Using the 160,000 per-query top-1 results of \toolc with and without synergy re-ranking (underlying \T~\ref{tab:rq5}),
a paired $t$-test confirms that the synergy gain is statistically
significant ($t=27.0$, $p=5.2 \times 10^{-161}$).
These results indicate that instruction alignment is complementary to hard negative mining: hard negatives sharpen the function-level decision boundary, while instruction alignment supplies fine-grained supervision that function-level triplets alone can hardly capture.
}

\subsection{Case Study}
\label{sec:evaluation:casestudy}

\begin{figure}
\centering
\includegraphics[width=\linewidth]{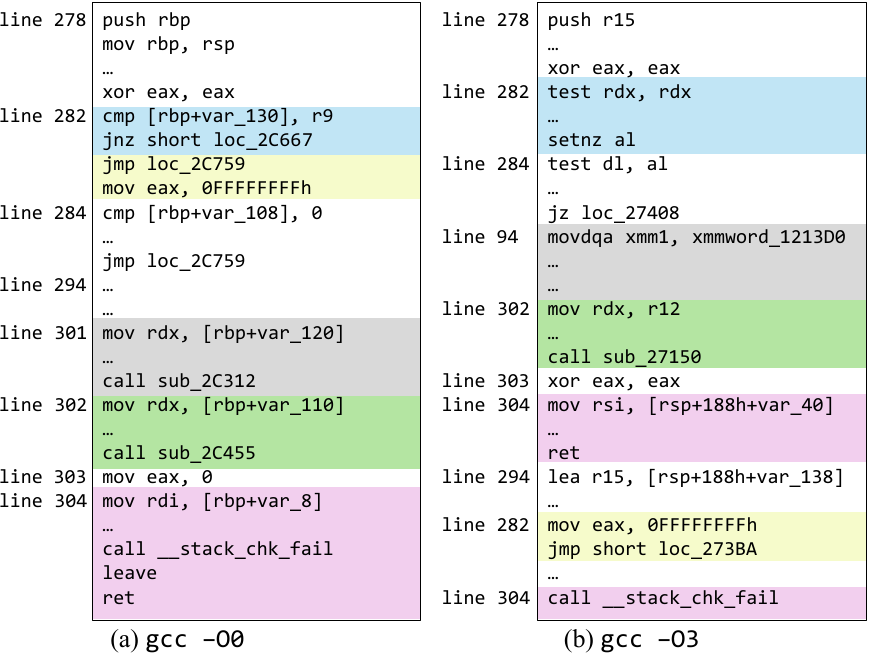}
\caption{Simplified assembly functions of \texttt{blake2b}. The line numbers denote their corresponding source lines.}
\label{fig:case-study}
\end{figure}

\begin{figure}[t]
    \centering
    \includegraphics[width=0.85\linewidth]{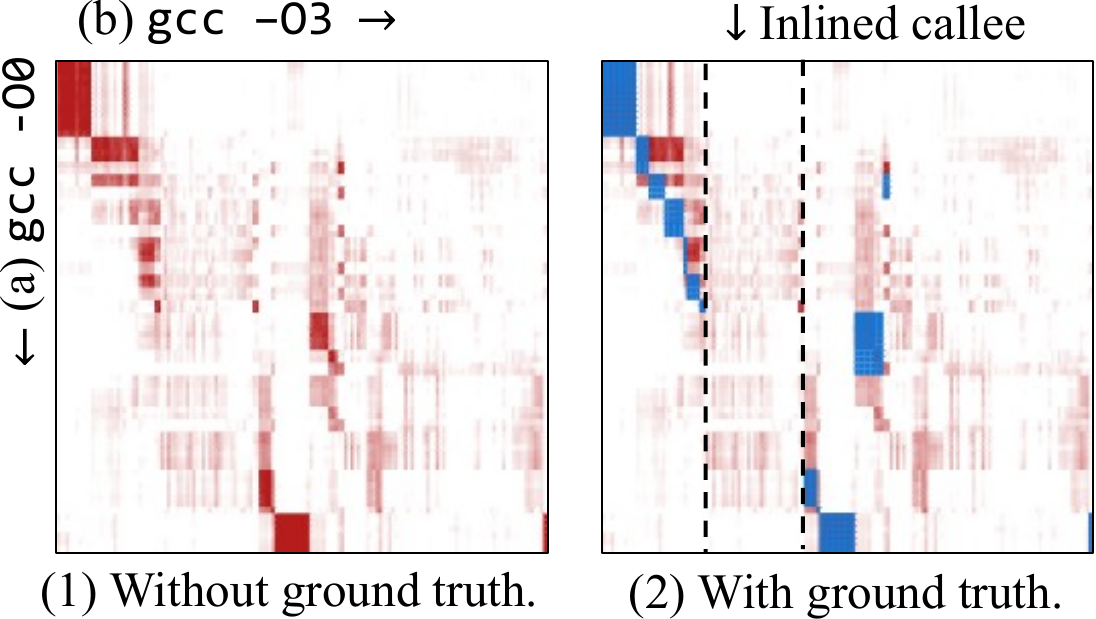}
    \caption{Instruction alignment similarity heatmap of \F~\ref{fig:case-study}.
    A red cell $(i,j)$ of (1) denotes a high similarity ($\mathbf{S}_{i,j}$) defined in \E~\ref{eq:insn-cos};
    a blue cell $(i,j)$ of (2) denotes $\mathbf{M}_{i,j}=1$ (\S\ref{sec:design:infonce}).
    The columns between two black dotted vertical lines of (2) have no blue cells, indicating that those instructions belong to inlined callees (i.e., \colorbox{lightgray}{line 94} of \F~\ref{fig:case-study}(b)).
    }
    \label{fig:heatmap}
\end{figure}

\setlength{\fboxrule}{0.3pt}
\setlength{\fboxsep}{0pt}

We present concrete examples illustrating how instruction alignment provides interpretable evidence for BCSA.
\F~\ref{fig:case-study} shows the simplified assembly of \texttt{blake2b} from Coreutils compiled by \texttt{gcc -O0} and \texttt{gcc -O3}, which exhibit significantly different instruction sequences and control flow structures.

\smallskip
\parh{Instruction Splitting Under Optimization.}
A single source line can be compiled into multiple non-contiguous instructions.
For instance, source line 282 produces four instructions marked by \colorbox{lightblue}{blue} and \colorbox{lightyellow}{yellow} in \F~\ref{fig:case-study}; under \texttt{-O3}, these are scattered across distant locations.
\toolj nonetheless assigns close embeddings to these instructions across the two variants, demonstrating resilience to code fragmentation.
Similarly, \colorbox{lightpurple}{line 304} is split in \F~\ref{fig:case-study}(b), with a compiler-inserted \texttt{call} interleaved between them.\looseness=-1

\smallskip
\parh{Robustness to Function Inlining.}
Function inlining poses a prominent challenge for BCSA~\cite{jia2024cross,functioninlining}, as the inlined callee's semantics are absent in the non-inlined variant.
In \F~\ref{fig:case-study}, the function called at \colorbox{lightgray}{line 301} in (a) is inlined into (b) (\colorbox{lightgray}{line 94}).
As shown in \F~\ref{fig:heatmap}, the model correctly assigns low similarity to instructions without ground-truth correspondences (nearly white cells), while the code marked by \colorbox{lightgreen}{green} in (a) achieves high similarity (around 0.7) with its counterparts in (b).
Comparing the predicted heatmap~(1) with the ground truth matrix~(2), the high-similarity regions closely match the true correspondences, confirming that the model identifies instructions from the same source line even in the presence of inlining.\looseness=-1

{
\smallskip
\parh{Alignment Failures.}
To understand the limitations of instruction alignment, we manually analyze all 216 BCSA failures of RQ5's best model on the validation set, i.e., query
functions (\texttt{gcc -O0}) whose top-1 \texttt{-O3} match is incorrect, and group them into four causes.\looseness=-1

\smallskip
\noindent\underline{Cause 1: Semantics-Deprived Wrapper Queries (162/216).}
The dominant cause is over-simple query functions.
At \texttt{-O0}, these functions are thin wrappers whose observable code merely
prepares the calling context (e.g., marshalling arguments) before delegating the real work to a callee.
The callee's semantics are invisible in the query, so the argument-preparation instructions carry little discriminative signal.
Consequently, the query aligns to many similar wrappers: its similarity matrix is dominated by high-similarity cells over shared boilerplate.
For instance, the following assembly code for \texttt{xcharalloc} is a wrapper that forwards its argument to \texttt{xmalloc}:\looseness=-1
}
\begin{lstlisting}[numbers=none]
push  rbp
mov   rbp, rsp
mov   [rbp+var_8], rdi  ; marshal argument
mov   rdi, [rbp+var_8]
call  sub_106436        ; invisible semantics (i.e., xmalloc)
leave
retn
\end{lstlisting}

\begin{figure}[t]
\centering
    \includegraphics[width=0.85\linewidth]{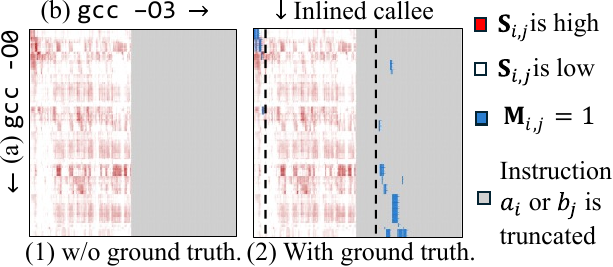}
    \caption{{Instruction alignment similarity heatmap of \texttt{cut\_file}. (a)$\downarrow$ $\times$ (b)$\rightarrow$ denotes $102 \times 465$ instruction alignment similarities ($S_{i,j}$).}}
    \label{fig:failure-cause2}
\end{figure}

{
\noindent\underline{Cause 2: Context-Window Truncation (14/216).}
When the optimized counterpart inlines many callees, its instruction sequence exceeds the model's input length (1,024 tokens) and is truncated, so the discriminative instructions never receive embeddings.
\F~\ref{fig:failure-cause2} shows this for \texttt{cut\_file}, aligning its $102$ \texttt{-O0} instructions against $465$ \texttt{-O3} instructions.
The columns between the two dashed lines in~(2) are inlined callees, which have no counterpart in the \texttt{-O0} query (no blue cells).
By consuming the context window, they push \texttt{cut\_file}'s own body past the token limit, truncating it (gray columns).
Many \texttt{-O0} instructions whose true counterparts fall in this truncated region (blue cells over gray) then find no high-similarity match, yielding near-zero row maxima that drag down the MAS.
This cause reflects an intrinsic input-length limit.
}

\smallskip
{
\noindent\underline{Cause 3: Near-Twin Siblings (20/216).}
The query and its true match share an almost identical body and differ only in a
single branch or one callee, yet the top-1 result is the sibling twin.
Because the differing callee's semantics are invisible, almost every query
instruction aligns with equally high similarity to both the twin and the true
match, so their similarity matrices are nearly identical;
An example is \texttt{xvprintf} vs. \texttt{xvfprintf}, which are
identical except for the delegated callee:
}
\begin{lstlisting}[numbers=none]
; xvprintf                    ; xvfprintf
call _vprintf                 call _vfprintf
... identical error handling (ferror, gettext, error) ...
\end{lstlisting}

{
\noindent\underline{Cause 4: Size, Constant, or Global-Only Differences (20/216).}
These memory-manipulation-dominated functions share the similar structure and differ only in memory sizes, magic constants, or the referenced global.
We find that corresponding instructions still earn high per-instruction similarity, saturating the similarity matrix and driving MAS close to one for both candidates; distinguishing them requires operand-level reasoning that the embeddings do not capture.
For instance, \texttt{md5\_init\_ctx} and \texttt{sha224\_init\_ctx} write the
same store pattern and differ only in the initialization constants and structure
size; under \texttt{-O3} both collapse to a vectorized load of an
initialization-vector constant:
}
\begin{lstlisting}[numbers=none]
; md5_init_ctx (-O3)          ; sha224_init_ctx (-O3)
movdqa xmm0, <ptr_A>          movdqa xmm0, <ptr_B>
movups [rdi], xmm0            movups [rdi], xmm0
mov    [rdi+14h], 0           movdqa xmm0, <ptr_C>
mov    [rdi+18h], 0           movups [rdi+10h], xmm0
\end{lstlisting}

\smallskip
{
\noindent\underline{Implication.}
Causes~1 and~2 stem from information absent from the query (invisible callee semantics or dropped code), and causes 3 and~4 stem from the model's inability to reason about subtle differences in operands.
Causes 1, 3, 4 are alignment false positives (high MAS on wrong candidates), while cause 2 often results in low MAS on the true match.

}

\begin{figure}
    \centering
    \begin{minipage}{0.9\linewidth}
\begin{lstlisting}[numbers=left]
cwrite(n_out == 0,hold,n_hold);
n_out += n_hold;
if (n_hold > bufsize) // Deleted in patch
  hold = xirealloc(hold,bufsize); // Deleted in patch
n_hold = 0;
hold_size = bufsize; // Deleted in patch
\end{lstlisting}
    \end{minipage}
\caption{The code snippet of vulnerable function \texttt{line\_bytes\_split}.}
\label{fig:vuln-code}
\end{figure}

\begin{figure}
\centering
\includegraphics[width=0.80\linewidth]{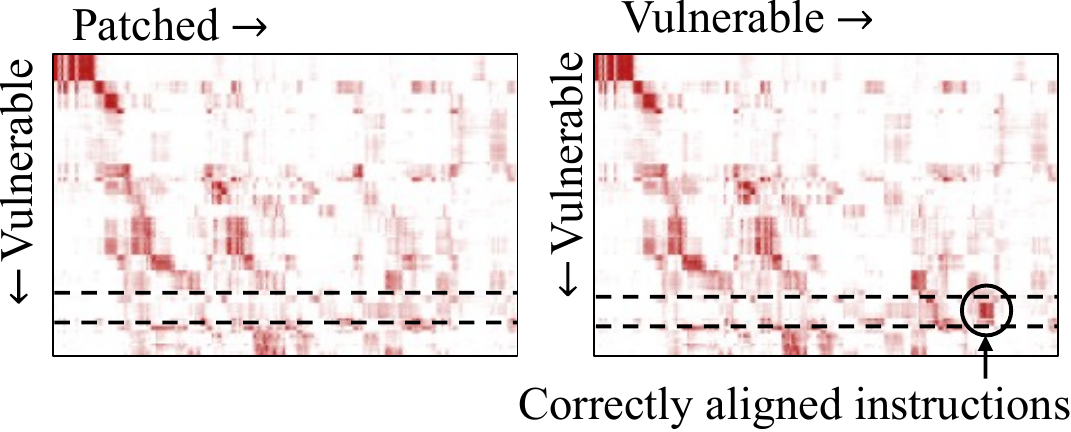}
\caption{Instruction alignment similarity heatmap for patch presence detection. The vulnerable function compiled with \texttt{gcc -O0}$\downarrow$ vs. patched and vulnerable functions compiled with \texttt{gcc -O3}$\rightarrow$.
The rows between two dotted horizontal lines correspond to assembly compiled from deleted lines of \F~\ref{fig:vuln-code}.
}
\label{fig:vuln-patch-heatmap}
\end{figure}

\smallskip
\parh{Patch Presence Detection.}
Beyond BCSA, instruction alignment may offer interpretable evidence for patch presence detection (PPD), which hinges on capturing subtle patch signatures~\cite{zhang2018precise,xu2020patch,he2022rapidpatch}.
Since a vulnerable function and its patched version typically share near-identical overall semantics, coarse-grained function-level representations struggle to tell them apart~\cite{zhang2018precise,bsa2bsca2024}.
We present a preliminary case to illustrate how fine-grained instruction alignment could help.

We study CVE-2024-0684~\cite{cve20240684}, affecting \texttt{line\_bytes\_split} in Coreutils (\F~\ref{fig:vuln-code}), whose patch removes lines 3--4 (line 6 is also deleted but optimized out under \texttt{-O3}).
We compile the vulnerable ver.~9.2 with \texttt{gcc -O0} and \texttt{gcc -O3}, and the patched ver.~9.5 with \texttt{gcc -O3}.
Queried with the \texttt{-O0} binary, \toolj ranks both the vulnerable and patched \texttt{-O3} variants at the top among over 2,700 Coreutils functions, and their function-level embeddings are too close to differentiate.
The instruction alignment heatmaps (\F~\ref{fig:vuln-patch-heatmap}), however, expose the difference: instructions from lines 3--4 of \F~\ref{fig:vuln-code} align to high-similarity counterparts in the vulnerable \texttt{-O3} variant but not in the patched one, where those source lines are absent.

{
This example illustrates the potential of instruction alignment for PPD, not a complete solution.
Realizing it must still overcome the intrinsic limitations discussed above, such as invisible callee semantics and limited input length.
For instance, the patch-relevant code may be truncated. We thus leave PPD application to future work.}\looseness=-1

 \section{Related Work}
\label{sec:related}

\parh{ML-Based Binary Code Similarity Analysis.}
ML-based BCSA methods learn semantic embeddings from assembly code for efficient similarity retrieval.
Representative approaches include graph neural networks~\cite{feng2016scalable,xu2017neural}, recurrent networks~\cite{massarelli2019safe,mikolov2010recurrent}, PV-DM with random walks~\cite{ding2019asm2vec,le2014distributed}, BERT-based pre-training~\cite{li2021palmtree,pei2022learning,devlin2018bert}, and Transformer architectures with jump-aware encodings~\cite{wang2022jtrans} or natural language supervision~\cite{wang2024clap}.
Other studies explored customized graph semantics~\cite{he2024hermessim,sem2vec2023,ncc} and comparison models~\cite{wang2024cebin,marcelli2022machine}.
These methods universally operate at the function level with symbol-based matching as the sole supervision signal~\cite{haq2021survey,marcelli2022machine} and contrastive learning as the dominant training paradigm.
Our work introduces instruction-level alignment as an auxiliary objective using multi-positive InfoNCE~\cite{khosla2020supervised}, complementing any existing embedding-based BCSA model.
Non-ML approaches~\cite{ming2017binsim,gitz2017,david2014tracelet,luo2014semantics,luo2017semantics,wang2022enhancing,wang2026vsim,manuel2014blanket,xu2023pem,wang2017imf,ye2026enhancing} are orthogonal to our method.

\smallskip
\parh{Fine-Grained Representations and Explainability.}
While most BCSA methods produce a single embedding per function, some work has explored finer granularities.
PalmTree~\cite{li2021palmtree} learns instruction embeddings via pre-training tasks, but these are not designed for cross-function instruction alignment;
we thus omit it in our evaluation.
DeepBinDiff~\cite{duan2020deep} generates basic-block-level embeddings for binary diffing, but often relies on semantically-irrelevant features like strings~\cite{wang2026vsim} and requires expensive pairwise computation.
In contrast, \tool learns instruction embeddings from instruction correspondences, which can be directly and efficiently aligned, providing fine-grained evidence of the semantic correspondence.\looseness=-1

\smallskip
{
\parh{LLM-Based Binary Code Analysis.}
Recent large-language-model-based (LLM-based) research has substantially expanded the scope of binary analysis.
Some studies directly decompile binary code~\cite{llm4decompile,ICLR2025ef283d62} or refine conventional decompiler outputs~\cite{llm4decompile,decllm2025,wong2023refining,degpt2024,binrag2026issta}, aiming to improve readability, recompilability, or semantic correctness.
Another line evaluates or adapts LLMs for binary comprehension tasks, such as function-name recovery~\cite{jiang2025beyond}, binary summarization~\cite{10795058}, similarity analysis~\cite{ICLR2025ef283d62}, and algorithm classification or multi-task binary reasoning~\cite{binmetric2025,llmsbt2026eurosys}.
These studies demonstrate the potential of LLMs for reverse engineering.
However, the explanations provided by them are often expressed as human-readable artifacts, such as decompiled code or summaries.
Such artifacts are cognitively useful, but readability alone does not guarantee that an explanation reflects a model's internal similarity decision or helps distinguish hard positive and negative pairs.\looseness=-1

In contrast, \tool focuses on binary representation learning for similarity retrieval. Its instruction-level alignments expose the fine-grained instruction correspondences between two binary functions, providing inspectable evidence for a similarity judgment rather than an external, post-hoc artifact, as the alignment is derived from the same instruction embeddings that form the function representation. We evaluate this signal not only qualitatively, but also through hard-negative discriminability and accuracy.
\tool is also related to explainable retrieval methods such as XSearch~\cite{xsearch2026issta}, which reformulates natural-language-to-code search as concept-to-code alignment. However, XSearch aligns query concepts with source-code statements, whereas \tool aligns assembly instructions across binary variants.}\looseness=-1

 \section{Conclusion}
\label{sec:conclusion}

We propose instruction alignment as a fine-grained supervision signal for binary code representation learning.
By deriving instruction-level correspondences from compiler debug information, we augment function-level contrastive training with an auxiliary alignment objective that improves both instruction-level and function-level embedding quality across diverse compiler configurations;
{
the resulting instruction-level alignment provides a more discriminative signal and inspectable evidence for similarity judgments.}

\section*{Acknowledgements}
This paper was supported in part by a grant from the Research Grants Council of the Hong Kong Special Administrative Region, China HKUST (No. C6004-25G) and an ITF grant under the contract ITS/161/24FP.

\section*{Data Availability Statement}
\label{sec:data-availability}
The code and data are available at \url{https://doi.org/10.5281/zenodo.19343892}, and we promise to maintain a public repository \url{https://github.com/whj0401/InsnAlign} for future research.
 
\bibliographystyle{ACM-Reference-Format}
\bibliography{bib/similarity,bib/sca,bib/machine-learning,
bib/other,bib/decompiler,bib/symbolic-execution,bib/analysis,bib/sw}

@inproceedings{jiang2025beyond,
  title={Beyond Classification: Inferring Function Names in Stripped Binaries via Domain Adapted LLMs},
  author={Jiang, Linxi and Jin, Xin and Lin, Zhiqiang},
  booktitle={Network and Distributed System Security (NDSS) Symposium},
  year={2025},
  doi={10.14722/ndss.2025.240797}
}

@inproceedings{xu2020patch,
  title={Patch based vulnerability matching for binary programs},
  author={Xu, Yifei and Xu, Zhengzi and Chen, Bihuan and Song, Fu and Liu, Yang and Liu, Ting},
  booktitle={Proceedings of the 29th ACM SIGSOFT International Symposium on Software Testing and Analysis},
  pages={376--387},
  year={2020},
  doi={10.1145/3395363.3397361}
}

@inproceedings{he2022rapidpatch,
  title={$\{$RapidPatch$\}$: firmware hotpatching for $\{$Real-Time$\}$ embedded devices},
  author={He, Yi and Zou, Zhenhua and Sun, Kun and Liu, Zhuotao and Xu, Ke and Wang, Qian and Shen, Chao and Wang, Zhi and Li, Qi},
  booktitle={31st USENIX Security Symposium (USENIX Security 22)},
  pages={2225--2242},
  year={2022}
}

@inproceedings{binmetric2025,
author = {Shang, Xiuwei and Chen, Guoqiang and Cheng, Shaoyin and Wu, Benlong and Hu, Li and Li, Gangyang and Zhang, Weiming and Yu, Nenghai},
title = {BinMetric: a comprehensive binary code analysis benchmark for large language models},
year = {2025},
isbn = {978-1-956792-06-5},
url = {https://doi.org/10.24963/ijcai.2025/858},
doi = {10.24963/ijcai.2025/858},
booktitle = {Proceedings of the Thirty-Fourth International Joint Conference on Artificial Intelligence},
articleno = {858},
numpages = {9},
location = {Montreal, Canada},
series = {IJCAI '25}
}

@INPROCEEDINGS {10795058,
author = { Shang, Xiuwei and Cheng, Shaoyin and Chen, Guoqiang and Zhang, Yanming and Hu, Li and Yu, Xiao and Li, Gangyang and Zhang, Weiming and Yu, Nenghai },
booktitle = { 2024 IEEE International Conference on Software Maintenance and Evolution (ICSME) },
title = {{ How Far Have We Gone in Binary Code Understanding Using Large Language Models }},
year = {2024},
volume = {},
ISSN = {},
pages = {1-12},
doi = {10.1109/ICSME58944.2024.00012},
url = {https://doi.ieeecomputersociety.org/10.1109/ICSME58944.2024.00012},
publisher = {IEEE Computer Society},
address = {Los Alamitos, CA, USA},
month =Oct}

@misc{xsearch2026issta,
      title={{XS}earch: Explainable Code Search via Concept-to-Code Alignment}, 
      author={Yiming Liu and Ruofan Liu and Yun Lin and Zicong Zhang and Weiyu Kong and Pengnian Qi and Xiao Cheng and Weinan Zhang and Qianxiang Wang and Linpeng Huang},
      year={2026},
      eprint={2605.16046},
      archivePrefix={arXiv},
      primaryClass={cs.SE},
      url={https://arxiv.org/abs/2605.16046}, 
}

@INPROCEEDINGS{binrag2026issta,
 author = {Wai Kin, Wong and Daoyuan, Wu and Zhibo, Liu and Huaijin, Wang and Zongjie, Li and Shuai, Wang},
 title = {{BinRAG}: An RAG-Based Decompilation Framework Fusing Name Prediction and Calling Context},
 booktitle = {Proceedings of the 2026 International Symposium on Software Testing and Analysis},
 series = {ISSTA '26},
 year = {2026},
 doi = {10.1145/3832245}
}

@INPROCEEDINGS{unseendelta2026issta,
 author = {Zhibo, Liu and Huaijin, Wang and Shuai, Wang},
 title = {The Unseen Delta: Characterizing the Compiler Optimization Landscape via Top-Down Differential Analysis},
 booktitle = {Proceedings of the 2026 International Symposium on Software Testing and Analysis},
 series = {ISSTA '26},
 year = {2026},
 doi={10.1145/3832164}
}

@inproceedings{llmsbt2026eurosys,
author = {Liu, Zhibo and Wang, Huaijin and Wong, Wai Kin and Wu, Daoyuan and Wang, Shuai},
title = {No More Translation at Runtime: LLM-Empowered Static Binary Translation},
year = {2026},
isbn = {9798400722127},
publisher = {Association for Computing Machinery},
address = {New York, NY, USA},
url = {https://doi.org/10.1145/3767295.3803600},
doi = {10.1145/3767295.3803600},
booktitle = {Proceedings of the 21st European Conference on Computer Systems},
pages = {1023--1040},
numpages = {18},
location = {McEwan Hall/The University of Edinburgh, Edinburgh, Scotland UK},
series = {EUROSYS '26}
}

@misc{ida,
  title={{IDA Pro: Powerful Disassembler, Decompiler \& Debugger}},
  author={Hex-Rays, SA},
  howpublished = {\url{https://hex-rays.com/ida-pro}},
  year = {2026}
}

@article{decllm2025,
author = {Wong, Wai Kin and Wu, Daoyuan and Wang, Huaijin and Li, Zongjie and Liu, Zhibo and Wang, Shuai and Tang, Qiyi and Nie, Sen and Wu, Shi},
title = {{DecLLM}: LLM-Augmented Recompilable Decompilation for Enabling Programmatic Use of Decompiled Code},
year = {2025},
publisher = {Association for Computing Machinery},
address = {New York, NY, USA},
volume = {2},
number = {ISSTA},
journal = {Proc. ACM Softw. Eng.},
month = jun,
articleno = {ISSTA081},
numpages = {24},
doi={10.1145/3728958}
}

@article{wong2023refining,
  title={Refining decompiled c code with large language models},
  author={Wong, Wai Kin and Wang, Huaijin and Li, Zongjie and Liu, Zhibo and Wang, Shuai and Tang, Qiyi and Nie, Sen and Wu, Shi},
  journal={arXiv preprint arXiv:2310.06530},
  year={2023}
}

@inproceedings{liu2022sok,
  title={{SoK}: Demystifying binary lifters through the lens of downstream applications},
  author={Liu, Zhibo and Yuan, Yuanyuan and Wang, Shuai and Bao, Yuyan},
  booktitle={2022 IEEE Symposium on Security and Privacy (SP)},
  pages={1100--1119},
  year={2022},
  organization={IEEE},
  doi={10.1109/SP46214.2022.9833799}
}

@inproceedings{llm4decompile,
    title = "{LLM}4{D}ecompile: Decompiling Binary Code with Large Language Models",
    author = "Tan, Hanzhuo  and
      Luo, Qi  and
      Li, Jing  and
      Zhang, Yuqun",
    editor = "Al-Onaizan, Yaser  and
      Bansal, Mohit  and
      Chen, Yun-Nung",
    booktitle = "Proceedings of the 2024 Conference on Empirical Methods in Natural Language Processing",
    month = nov,
    year = "2024",
    address = "Miami, Florida, USA",
    publisher = "Association for Computational Linguistics",
    doi = "10.18653/v1/2024.emnlp-main.203",
    pages = "3473--3487"
}

@inproceedings{ICLR2025ef283d62,
 author = {Jiang, Nan and Wang, Chengxiao and Liu, Kevin and Xu, Xiangzhe and Tan, Lin and Zhang, Xiangyu and Babkin, Petr},
 booktitle = {International Conference on Learning Representations},
 pages = {95905--95926},
 title = {Nova: Generative Language Models for Assembly Code with Hierarchical Attention and Contrastive Learning},
 url = {https://proceedings.iclr.cc/paper_files/paper/2025/file/ef283d62b4bce30854a8d4827f331229-Paper-Conference.pdf},
 year = {2025}
}

@inproceedings{degpt2024,
  title={De{GPT}: Optimizing Decompiler Output with LLM},
  author={Peiwei, Hu and Ruigang, Liang and Kai, Chen},
	booktitle = {NDSS},
  year={2024},
  doi={10.14722/ndss.2024.24401}
}

@article{devlin2018bert,
  title={{BERT}: Pre-training of deep bidirectional transformers for language understanding},
  author={Devlin, Jacob},
  journal={arXiv preprint arXiv:1810.04805},
  year={2018}
}

@inproceedings{mikolov2010recurrent,
  title={Recurrent neural network based language model},
  author={Mikolov, Tom{\'a}{\v{s}} and Karafi{\'a}t, Martin and Burget, Luk{\'a}{\v{s}} and {\v{C}}ernock{\`y}, Jan and Khudanpur, Sanjeev},
  booktitle={Eleventh annual conference of the international speech communication association},
  year={2010}
}

@inproceedings{le2014distributed,
author = {Le, Quoc and Mikolov, Tomas},
title = {Distributed representations of sentences and documents},
year = {2014},
publisher = {JMLR.org},
booktitle = {Proceedings of the 31st International Conference on International Conference on Machine Learning - Volume 32},
pages = {II–1188–II–1196},
location = {Beijing, China},
series = {ICML'14}
}

@article{oord2018representation,
  title={Representation learning with contrastive predictive coding},
  author={Oord, Aaron van den and Li, Yazhe and Vinyals, Oriol},
  journal={arXiv preprint arXiv:1807.03748},
  year={2018}
}

@article{bradley1997use,
title = {The use of the area under the ROC curve in the evaluation of machine learning algorithms},
journal = {Pattern Recognition},
volume = {30},
number = {7},
pages = {1145-1159},
year = {1997},
issn = {0031-3203},
doi = {https://doi.org/10.1016/S0031-3203(96)00142-2},
author = {Andrew P. Bradley}
}

@article{hand2023notes,
  title={Notes on the H-measure of classifier performance},
  author={Hand, David J and Anagnostopoulos, Christoforos},
  journal={Advances in Data Analysis and Classification},
  volume={17},
  number={1},
  pages={109--124},
  year={2023},
  publisher={Springer},
  doi={10.1007/s11634-021-00490-3}
}

@article{saito2015precision,
  title={The precision-recall plot is more informative than the ROC plot when evaluating binary classifiers on imbalanced datasets},
  author={Saito, Takaya and Rehmsmeier, Marc},
  journal={PloS one},
  volume={10},
  number={3},
  pages={e0118432},
  year={2015},
  publisher={Public Library of Science}
}

@book{cohen2013statistical,
  title={Statistical power analysis for the behavioral sciences},
  author={Cohen, Jacob},
  year={2013},
  publisher={routledge},
  doi={10.4324/9780203771587}
}

@inproceedings{chen2018gradnorm,
  title = 	 {{G}rad{N}orm: Gradient Normalization for Adaptive Loss Balancing in Deep Multitask Networks},
  author =       {Chen, Zhao and Badrinarayanan, Vijay and Lee, Chen-Yu and Rabinovich, Andrew},
  booktitle = 	 {Proceedings of the 35th International Conference on Machine Learning},
  pages = 	 {794--803},
  year = 	 {2018},
  editor = 	 {Dy, Jennifer and Krause, Andreas},
  volume = 	 {80},
  series = 	 {Proceedings of Machine Learning Research},
  month = 	 {10--15 Jul},
  publisher =    {PMLR},
  url = 	 {https://proceedings.mlr.press/v80/chen18a.html},
}

@article{khosla2020supervised,
 author = {Khosla, Prannay and Teterwak, Piotr and Wang, Chen and Sarna, Aaron and Tian, Yonglong and Isola, Phillip and Maschinot, Aaron and Liu, Ce and Krishnan, Dilip},
 booktitle = {Advances in Neural Information Processing Systems},
 pages = {18661--18673},
 publisher = {Curran Associates, Inc.},
 title = {Supervised Contrastive Learning},
 url = {https://proceedings.neurips.cc/paper_files/paper/2020/file/d89a66c7c80a29b1bdbab0f2a1a94af8-Paper.pdf},
 year = {2020}
}

@inproceedings{schroff2015facenet,
  title={Facenet: A unified embedding for face recognition and clustering},
  author={Schroff, Florian and Kalenichenko, Dmitry and Philbin, James},
  booktitle={Proceedings of the IEEE conference on computer vision and pattern recognition},
  pages={815--823},
  year={2015},
  doi={10.1109/CVPR.2015.7298682}
}

@article{liu2019roberta,
  title={Roberta: A robustly optimized bert pretraining approach},
  author={Liu, Yinhan and Ott, Myle and Goyal, Naman and Du, Jingfei and Joshi, Mandar and Chen, Danqi and Levy, Omer and Lewis, Mike and Zettlemoyer, Luke and Stoyanov, Veselin},
  journal={arXiv preprint arXiv:1907.11692},
  year={2019}
}

@misc{robinson2021contrastive,
      title={Contrastive Learning with Hard Negative Samples}, 
      author={Joshua Robinson and Ching-Yao Chuang and Suvrit Sra and Stefanie Jegelka},
      year={2021},
      eprint={2010.04592},
      archivePrefix={arXiv},
      primaryClass={cs.LG},
      url={https://arxiv.org/abs/2010.04592}, 
}

@manual{dwarf5,
  title        = {DWARF Debugging Information Format},
  author       = {{DWARF Debugging Information Format Committee}},
  version      = {Version 5},
  year         = {2017},
  url          = {https://dwarfstd.org/doc/DWARF5.pdf},
}

@manual{cve20240684,
  author = {Paul Eggert},
  title = {Coreutils fix for CVE-2024-0684.},
  year = {2026},
  url = {{https://github.com/coreutils/coreutils/commit/c4c5ed8f4e9cd55a12966d4f520e3a13101637d9}}
}

@inproceedings{yu2020codecmr,
 author = {Yu, Zeping and Zheng, Wenxin and Wang, Jiaqi and Tang, Qiyi and Nie, Sen and Wu, Shi},
 booktitle = {Advances in Neural Information Processing Systems},
 pages = {3872--3883},
 publisher = {Curran Associates, Inc.},
 title = {{CodeCMR}: Cross-Modal Retrieval For Function-Level Binary Source Code Matching},
 volume = {33},
 year = {2020}
}

@inproceedings{jiang2024binaryai,
author = {Jiang, Ling and An, Junwen and Huang, Huihui and Tang, Qiyi and Nie, Sen and Wu, Shi and Zhang, Yuqun},
title = {{BinaryAI}: Binary Software Composition Analysis via Intelligent Binary Source Code Matching},
year = {2024},
doi = {10.1145/3597503.3639100},
booktitle = {Proceedings of the IEEE/ACM 46th International Conference on Software Engineering},
articleno = {224},
numpages = {13},
series = {ICSE '24}
}

@INPROCEEDINGS{bsa2bsca2024,
  author={Wang, Huaijin and Liu, Zhibo and Wang, Shuai and Wang, Ying and Tang, Qiyi and Nie, Sen and Wu, Shi},
  booktitle={2024 IEEE 9th European Symposium on Security and Privacy}, 
  title={Are We There Yet? Filling the Gap Between Binary Similarity Analysis and Binary Software Composition Analysis}, 
  year={2024},
  volume={},
  number={},
  pages={506-523},
  doi={10.1109/EuroSP60621.2024.00034}}

@inproceedings{safesca2025,
author = {Wang, Huaijin and Liu, Zhibo and Dai, Yanbo and Wang, Shuai and Tang, Qiyi and Nie, Sen and Wu, Shi},
title = {Preserving Privacy in Software Composition Analysis: A Study of Technical Solutions and Enhancements},
year = {2025},
booktitle = {Proceedings of the IEEE/ACM 47th International Conference on Software Engineering (ICSE)},
pages = {2329--2341},
numpages = {13},
doi={10.1109/ICSE55347.2025.00055}
}

@inproceedings{luo2014semantics,
  title={Semantics-based obfuscation-resilient binary code similarity comparison with applications to software plagiarism detection},
  author={Luo, Lannan and Ming, Jiang and Wu, Dinghao and Liu, Peng and Zhu, Sencun},
  booktitle={Proceedings of the 22nd ACM SIGSOFT international symposium on foundations of software engineering},
  pages={389--400},
  year={2014},
  doi = {10.1145/2635868.2635900},
}

@article{luo2017semantics,
  title={Semantics-based obfuscation-resilient binary code similarity comparison with applications to software and algorithm plagiarism detection},
  author={Luo, Lannan and Ming, Jiang and Wu, Dinghao and Liu, Peng and Zhu, Sencun},
  journal={IEEE Transactions on Software Engineering},
  volume={43},
  number={12},
  pages={1157--1177},
  year={2017},
  publisher={IEEE},
  doi = {10.1109/TSE.2017.2655046},
}

@inproceedings{manuel2014blanket,
author = {Manuel Egele and Maverick Woo and Peter Chapman and David Brumley},
title = {Blanket Execution: Dynamic Similarity Testing for Program Binaries and Components},
booktitle = {Proceedings of the 23rd USENIX Security Symposium},
year = {2014},
month = Aug,
pages = {303-317},
publisher = {USENIX Association},
}

@inproceedings{david2014tracelet,
author = {David, Yaniv and Yahav, Eran},
title = {Tracelet-based code search in executables},
year = {2014},
isbn = {9781450327848},
address = {New York, NY, USA},
doi = {10.1145/2594291.2594343},
booktitle = {Proceedings of the 35th ACM SIGPLAN Conference on Programming Language Design and Implementation},
pages = {349--360},
numpages = {12},
location = {Edinburgh, United Kingdom},
series = {PLDI '14}
}

@inproceedings{ming2017binsim,
  title={Binsim: Trace-based semantic binary diffing via system call sliced segment equivalence checking},
  author={Ming, Jiang and Xu, Dongpeng and Jiang, Yufei and Wu, Dinghao},
  booktitle={Proceedings of the 26th USENIX Security Symposium},
  year={2017}
}

@inproceedings{chandramohan2016bingo,
author = {Chandramohan, Mahinthan and Xue, Yinxing and Xu, Zhengzi and Liu, Yang and Cho, Chia Yuan and Tan, Hee Beng Kuan},
title = {BinGo: cross-architecture cross-OS binary search},
year = {2016},
isbn = {9781450342186},
publisher = {Association for Computing Machinery},
address = {New York, NY, USA},
doi = {10.1145/2950290.2950350},
booktitle = {Proceedings of the 2016 24th ACM SIGSOFT International Symposium on Foundations of Software Engineering},
pages = {678--689},
numpages = {12},
location = {Seattle, WA, USA},
series = {FSE 2016}
}

@inproceedings{wang2022jtrans,
author = {Wang, Hao and Qu, Wenjie and Katz, Gilad and Zhu, Wenyu and Gao, Zeyu and Qiu, Han and Zhuge, Jianwei and Zhang, Chao},
title = {{jTrans}: jump-aware transformer for binary code similarity detection},
year = {2022},
booktitle = {Proceedings of the 31st ACM SIGSOFT International Symposium on Software Testing and Analysis},
pages = {1-13},
numpages = {13},
series = {ISSTA},
doi={10.1145/3533767.3534367}
}

@inproceedings{ncc,
 author = {Ben-Nun, Tal and Jakobovits, Alice Shoshana and Hoefler, Torsten},
 booktitle = {Advances in Neural Information Processing Systems},
 editor = {S. Bengio and H. Wallach and H. Larochelle and K. Grauman and N. Cesa-Bianchi and R. Garnett},
 pages = {},
 publisher = {Curran Associates, Inc.},
 title = {Neural Code Comprehension: A Learnable Representation of Code Semantics},
 url = {https://proceedings.neurips.cc/paper_files/paper/2018/file/17c3433fecc21b57000debdf7ad5c930-Paper.pdf},
 volume = {31},
 year = {2018}
}

@inproceedings{duan2020deep,
  title={{DeepBinDiff}: Learning Program-Wide Code Representations for Binary Diffing},
  author={Duan, Yue and Li, Xuezixiang and Wang, Jinghan and Yin, Heng},
  booktitle={Network and Distributed Systems Security (NDSS) Symposium},
  year={2020},
  doi={10.14722/ndss.2020.24311}
}

@inproceedings{gitz2017,
author = {David, Yaniv and Partush, Nimrod and Yahav, Eran},
title = {Similarity of binaries through re-optimization},
year = {2017},
isbn = {9781450349888},
address = {New York, NY, USA},
doi = {10.1145/3062341.3062387},
booktitle = {Proceedings of the 38th ACM SIGPLAN Conference on Programming Language Design and Implementation},
pages = {79--94},
numpages = {16},
location = {Barcelona, Spain},
series = {PLDI 2017}
}

@article{haq2021survey,
author = {Haq, Irfan Ul and Caballero, Juan},
title = {A Survey of Binary Code Similarity},
year = {2021},
issue_date = {April 2022},
publisher = {Association for Computing Machinery},
address = {New York, NY, USA},
volume = {54},
number = {3},
issn = {0360-0300},
url = {https://doi.org/10.1145/3446371},
doi = {10.1145/3446371},
journal = {ACM Comput. Surv.},
month = apr,
articleno = {51},
numpages = {38}
}

@inproceedings{massarelli2019safe,
  title={{SAFE}: Self-attentive function embeddings for binary similarity},
  author={Massarelli, Luca and Luna, Giuseppe Antonio Di and Petroni, Fabio and Baldoni, Roberto and Querzoni, Leonardo},
  booktitle={International Conference on Detection of Intrusions and Malware, and Vulnerability Assessment},
  pages={309--329},
  year={2019},
  organization={Springer},
  doi={10.1007/978-3-030-22038-9_15}
}

@inproceedings{wang2017imf,
  title={In-memory fuzzing for binary code similarity analysis},
  author={Wang, Shuai and Wu, Dinghao},
  booktitle={2017 32nd IEEE/ACM International Conference on Automated Software Engineering (ASE)},
  pages={319--330},
  year={2017},
  doi={10.5555/3155562.3155606}
}

@article{kim2022revisiting,
  author={Kim, Dongkwan and Kim, Eunsoo and Cha, Sang Kil and Son, Sooel and Kim, Yongdae},
  journal={IEEE Transactions on Software Engineering}, 
  title={Revisiting Binary Code Similarity Analysis Using Interpretable Feature Engineering and Lessons Learned}, 
  year={2023},
  volume={49},
  number={4},
  pages={1661-1682},
  doi={10.1109/TSE.2022.3187689}
}

@article{functioninlining,
  title={1-to-1 or 1-to-n? Investigating the Effect of Function Inlining on Binary Similarity Analysis},
  author={Jia, Ang and Fan, Ming and Jin, Wuxia and Xu, Xi and Zhou, Zhaohui and Tang, Qiyi and Nie, Sen and Wu, Shi and Liu, Ting},
  journal={ACM Transactions on Software Engineering and Methodology},
  volume={32},
  number={4},
  pages={1--26},
  year={2023},
  publisher={ACM New York, NY, USA}
}

@inproceedings{marcelli2022machine,
  title={How machine learning is solving the binary function similarity problem},
  author={Marcelli, Andrea and Graziano, Mariano and Ugarte-Pedrero, Xabier and Fratantonio, Yanick and Mansouri, Mohamad and Balzarotti, Davide},
  booktitle={31st USENIX Security Symposium (USENIX Security 22)},
  pages={2099--2116},
  year={2022}
}

@article{wang2022enhancing,
  title={Enhancing {DNN}-based binary code function search with low-cost equivalence checking},
  author={Wang, Huaijin and Ma, Pingchuan and Yuan, Yuanyuan and Liu, Zhibo and Wang, Shuai and Tang, Qiyi and Nie, Sen and Wu, Shi},
  journal={IEEE Transactions on Software Engineering},
  volume={49},
  number={1},
  pages={226--250},
  year={2022},
  publisher={IEEE},
  doi={10.1109/TSE.2022.3149240}
}

@article{sem2vec2023,
author = {Wang, Huaijin and Ma, Pingchuan and Wang, Shuai and Tang, Qiyi and Nie, Sen and Wu, Shi},
title = {sem2vec: Semantics-aware Assembly Tracelet Embedding},
year = {2023},
issue_date = {July 2023},
publisher = {Association for Computing Machinery},
address = {New York, NY, USA},
volume = {32},
number = {4},
issn = {1049-331X},
journal = {ACM Transactions on Software Engineering and Methodology},
month = may,
articleno = {90},
numpages = {34},
doi={10.1145/3569933}
}

@inproceedings{wang2024cebin,
author = {Wang, Hao and Gao, Zeyu and Zhang, Chao and Sun, Mingyang and Zhou, Yuchen and Qiu, Han and Xiao, Xi},
title = {{CEBin}: A Cost-Effective Framework for Large-Scale Binary Code Similarity Detection},
year = {2024},
booktitle = {Proceedings of the 33rd ACM SIGSOFT International Symposium on Software Testing and Analysis},
pages = {149--161},
numpages = {13},
series = {ISSTA},
doi={10.1145/3650212.3652117}
}

@inproceedings{wong2024binaug,
  title={{BinAug}: Enhancing Binary Similarity Analysis with Low-Cost Input Repairing},
  author={Wong, Wai Kin and Wang, Huaijin and Li, Zongjie and Wang, Shuai},
  booktitle={Proceedings of the 46th IEEE/ACM International Conference on Software Engineering},
  year={2024},
  articleno = {7},
  numpages = {13},
  location = {Lisbon, Portugal},
  doi = {10.1145/3597503.3623328},
}

@inproceedings{he2024hermessim,
  title={Code is not natural language: Unlock the power of semantics-oriented graph representation for binary code similarity detection},
  author={He, Haojie and Lin, Xingwei and Weng, Ziang and Zhao, Ruijie and Gan, Shuitao and Chen, Libo and Ji, Yuede and Wang, Jiashui and Xue, Zhi},
  booktitle={33rd USENIX Security Symposium (USENIX Security 24)},
  year={2024},
  pages = {1759-1776},
}

@inproceedings{wang2024clap,
  title={CLAP: Learning transferable binary code representations with natural language supervision},
  author={Wang, Hao and Gao, Zeyu and Zhang, Chao and Sha, Zihan and Sun, Mingyang and Zhou, Yuchen and Zhu, Wenyu and Sun, Wenju and Qiu, Han and Xiao, Xi},
  booktitle={Proceedings of the 33rd ACM SIGSOFT International Symposium on Software Testing and Analysis},
  pages={503--515},
  year={2024},
  doi = {10.1145/3650212.3652145},
}

@inproceedings{jia2024cross,
  title={Cross-inlining binary function similarity detection},
  author={Jia, Ang and Fan, Ming and Xu, Xi and Jin, Wuxia and Wang, Haijun and Liu, Ting},
  booktitle={Proceedings of the IEEE/ACM 46th International Conference on Software Engineering},
  pages={1--13},
  year={2024}
}

@inproceedings{feng2016scalable,
author = {Feng, Qian and Zhou, Rundong and Xu, Chengcheng and Cheng, Yao and Testa, Brian and Yin, Heng},
title = {Scalable Graph-based Bug Search for Firmware Images},
year = {2016},
doi = {10.1145/2976749.2978370},
booktitle = {Proceedings of the 2016 ACM SIGSAC Conference on Computer and Communications Security},
pages = {480--491},
numpages = {12},
series = {CCS '16}
}

@inproceedings{luo2023vulhawk,
  title={{VulHawk}: Cross-architecture Vulnerability Detection with Entropy-based Binary Code Search.},
  author={Luo, Zhenhao and Wang, Pengfei and Wang, Baosheng and Tang, Yong and Xie, Wei and Zhou, Xu and Liu, Danjun and Lu, Kai},
  booktitle={Network and Distributed Systems Security (NDSS) Symposium},
  year={2023},
  doi={10.14722/ndss.2023.24415}
}

@article{pei2022learning,
  author={Pei, Kexin and Xuan, Zhou and Yang, Junfeng and Jana, Suman and Ray, Baishakhi},
  journal={IEEE Transactions on Software Engineering}, 
  title={Learning Approximate Execution Semantics From Traces for Binary Function Similarity}, 
  year={2023},
  volume={49},
  number={4},
  pages={2776-2790},
  doi={10.1109/TSE.2022.3231621}
}

@inproceedings{xu2023pem,
author = {Xu, Xiangzhe and Xuan, Zhou and Feng, Shiwei and Cheng, Siyuan and Ye, Yapeng and Shi, Qingkai and Tao, Guanhong and Yu, Le and Zhang, Zhuo and Zhang, Xiangyu},
title = {{PEM}: Representing Binary Program Semantics for Similarity Analysis via a Probabilistic Execution Model},
year = {2023},
booktitle = {Proceedings of the 31st ACM Joint European Software Engineering Conference and Symposium on the Foundations of Software Engineering},
pages = {401--412},
numpages = {12},
doi={10.1145/3611643.3616301}
}

@inproceedings{ding2019asm2vec,
  title={Asm2vec: Boosting static representation robustness for binary clone search against code obfuscation and compiler optimization},
  author={Ding, Steven HH and Fung, Benjamin CM and Charland, Philippe},
  booktitle={2019 ieee symposium on security and privacy},
  pages={472--489},
  year={2019},
  organization={IEEE},
  doi={10.1109/SP.2019.00003}
}

@inproceedings{xu2017neural,
  title={Neural network-based graph embedding for cross-platform binary code similarity detection},
  author={Xu, Xiaojun and Liu, Chang and Feng, Qian and Yin, Heng and Song, Le and Song, Dawn},
  booktitle={Proceedings of the 2017 ACM SIGSAC Conference on Computer and Communications Security},
  pages={363--376},
  year={2017},
  doi={10.1145/3133956.3134018}
}

@inproceedings{zuo2019neural, 
title={Neural Machine Translation Inspired Binary Code Similarity Comparison beyond Function Pairs}, 
author={Zuo, Fei and Li, Xiaopeng and Young, Patrick and Luo,Lannan and Zeng,Qiang and Zhang, Zhexin}, 
year={2019},
booktitle={Network and Distributed Systems Security (NDSS) Symposium},
doi={10.14722/ndss.2019.23492}
}

@inproceedings{li2022unleashing,
  title={Unleashing the power of compiler intermediate representation to enhance neural program embeddings},
  author={Li, Zongjie and Ma, Pingchuan and Wang, Huaijin and Wang, Shuai and Tang, Qiyi and Nie, Sen and Wu, Shi},
  booktitle={Proceedings of the 44th International Conference on Software Engineering},
  pages={2253--2265},
  year={2022},
  doi = {10.1145/3510003.3510217}
}

@inproceedings{wong2022deceiving,
  author={Wong, Wai Kin and Wang, Huaijin and Ma, Pingchuan and Wang, Shuai and Jiang, Mingyue and Chen, Tsong Yueh and Tang, Qiyi and Nie, Sen and Wu, Shi},
  booktitle={2022 IEEE International Conference on Software Maintenance and Evolution}, 
  title={Deceiving Deep Neural Networks-Based Binary Code Matching with Adversarial Programs}, 
  year={2022},
  pages={117-128},
  doi={10.1109/ICSME55016.2022.00019}
}

@inproceedings{ye2026enhancing, 
title={Enhancing Semantic-Aware Binary Diffing with High-Confidence Dynamic Instruction Alignment}, 
author={Ye, Chengfeng and Zhou, Anshunkang and Zhang, Charles}, 
year={2026},
booktitle={Network and Distributed Systems Security (NDSS) Symposium},
doi={10.14722/ndss.2026.240663}
}

@inproceedings{wang2026vsim, 
  title={{vSim}: Semantics-aware value extraction for efficient binary code similarity analysis}, 
  author={Wang, Huaijin and Lin, Zhiqiang}, 
  year={2026},
  booktitle={Network and Distributed Systems Security (NDSS) Symposium},
  doi={10.14722/ndss.2026.240213}
}

@article{liu2025keenhash,
  title={{KEENHash}: Hashing programs into function-aware embeddings for large-scale binary code similarity analysis},
  author={Liu, Zhijie and Tang, Qiyi and Nie, Sen and Wu, Shi and Zhang, Liang Feng and Tang, Yutian},
  journal={Proceedings of the ACM on Software Engineering},
  volume={2},
  number={ISSTA},
  pages={801--824},
  year={2025},
  doi = {10.1145/3728911}
}

@inproceedings{li2021palmtree,
  title={Palmtree: learning an assembly language model for instruction embedding},
  author={Li, Xuezixiang and Qu, Yu and Yin, Heng},
  booktitle={Proceedings of the 2021 ACM SIGSAC Conference on Computer and Communications Security},
  pages={3236--3251},
  year={2021},
  doi = {10.1145/3460120.3484587}
}

@inproceedings{yu2020order,
  title={Order matters: Semantic-aware neural networks for binary code similarity detection},
  author={Yu, Zeping and Cao, Rui and Tang, Qiyi and Nie, Sen and Huang, Junzhou and Wu, Shi},
  booktitle={Proceedings of the AAAI conference on artificial intelligence},
  volume={34},
  number={01},
  pages={1145--1152},
  year={2020},
  doi={10.1609/aaai.v34i01.5466}
}

@article{lu2024dtd,
  title={Dtd: Comprehensive and scalable testing for debuggers},
  author={Lu, Hongyi and Liu, Zhibo and Wang, Shuai and Zhang, Fengwei},
  journal={Proceedings of the ACM on Software Engineering},
  volume={1},
  number={FSE},
  pages={1172--1193},
  year={2024},
  doi = {10.1145/3643779},
}

@ARTICLE{rlobf2020,
  author={Wang, Huaijin and Wang, Shuai and Xu, Dongpeng and Zhang, Xiangyu and Liu, Xiao},
  journal={IEEE Transactions on Dependable and Secure Computing}, 
  title={Generating Effective Software Obfuscation Sequences With Reinforcement Learning}, 
  year={2022},
  volume={19},
  number={3},
  pages={1900-1917},
  doi={10.1109/TDSC.2020.3041655}}

@inproceedings{zhang2018precise,
  title={Precise and accurate patch presence test for binaries},
  author={Zhang, Hang and Qian, Zhiyun},
  booktitle={27th USENIX Security Symposium (USENIX Security 18)},
  pages={887--902},
  year={2018}
}

\end{document}